\documentclass[12pt]{article}
\usepackage{hyperref}
\usepackage{graphicx}
\usepackage{cite}
\usepackage{amsmath,amssymb}
\usepackage{epstopdf}
\usepackage{lineno}
\usepackage{authblk}
\usepackage[margin=1.0in]{geometry}
\usepackage{tabularx}

\input epsf.tex    
\input epsf.def   

\input mydefs.sty

\begin{document}

\begin{titlepage}

\title{Status and Implications of BSM Searches at the LHC}
\date{}

\author[1]{Eva Halkiadakis}
\author[2]{George Redlinger}
\author[3]{David Shih}

\affil[1]{{\small Dept of Physics and
    Astronomy, Rutgers University, Piscataway, NJ 08854}}
\affil[2]{{\small Physics Dept, Brookhaven
    National Laboratory, Upton, NY 11973}}
\affil[3]{{\small NHETC, Dept of Physics and Astronomy, Rutgers University,
    Piscataway, NJ 08854}}

\maketitle
\thispagestyle{empty}

\vspace{0.4cm}

\begin{abstract}
The LHC has collided protons on protons at center-of-mass
energies of 7 and 8~TeV between 2010-2012, referred to as the Run I
period.  We review the current status of searches for new physics
beyond the Standard Model at the end of Run I by the ATLAS and CMS
experiments, limited to the 8~TeV search results published or submitted
for publication as of the end of February 2014. We discuss some of the implications of these
searches on the existence of TeV scale new physics, with a special
focus on two open questions: the hierarchy problem, and the nature of
dark matter.  Finally, we give an outlook for the future.

\end{abstract}

\end{titlepage}

\newpage

\section{INTRODUCTION}
\label{sec:Introduction}

The LHC is a proton-proton collider, designed to operate at a
center-of-mass energy of $\sqrt{s}=14$~TeV and to collect on the order
of 100-300 \ifb\ of data \cite{Evans:2011zzb}.  At the end of 2012, the two multi-purpose
LHC experiments (ATLAS \cite{Aad:2008zzm} and CMS
\cite{Chatrchyan:2008aa}) concluded what has come to be known as
``Run~I'' in which the LHC operated at $\sqrt{s}=7$ and 8~TeV,
collecting integrated luminosities of approximately 5~\ifb\ and
20~\ifb, respectively, at the two energies.  Even at these reduced
energies, the LHC has well explored the \TeV\ scale.

The physics program of ATLAS and CMS rests on two pillars:
\begin{itemize}
  \item Elucidating the mechanism of electroweak symmetry breaking (EWSB)
  \item Searching for physics Beyond the Standard Model (BSM)
\end{itemize}
With the discovery \cite{Aad:2012tfa, Chatrchyan:2012ufa} in July 2012
of what appears to be the Standard Model (SM)
Higgs Boson, the first part of the program has passed an important
milestone.  While much effort will still be devoted to the important
task of precisely measuring the properties of this SM-like Higgs
boson, there is now renewed attention to the second major goal of the
LHC.  The search for BSM physics has already been an active and robust
activity at the LHC. It is only expected to intensify at Run II when
the machine is upgraded to its design energy.

To a large extent, searches for new physics have been motivated by two
long-standing puzzles:

{\bf I. The Hierarchy Problem:} 
It was recognized long ago that the SM Higgs -- along with any other
elementary scalar particles -- suffers from what has come to be known
variously as the ``naturalness'', ``hierarchy'' or ``fine-tuning''
problem \cite{Weinberg:1975gm, Gildener:1976ai, Weinberg:1979bn,
  Susskind:1978ms,'tHooft:1979bh}. Today it is most common to describe the problem in
terms of the radiative corrections to the Higgs mass which depend
quadratically on the high-energy cutoff $\Lambda$ up to which the SM is
valid. For instance, one has
\begin{equation}
\label{eq:TeVscale}
\delta m_h^2 \sim {y_t^2\over 16\pi^2} \Lambda^2 ~,
\end{equation}
at one-loop in the SM from the coupling ($y_{t} \approx 1$) to the top
quark.  Requiring that this be of the same order as the Higgs mass
itself, one arrives to the conclusion that the cutoff scale, and
therefore the appearance of new physics, should be in the 1~TeV range.
Conversely, if the SM is valid all the way up to the Planck scale
($\Lambda \approx 10^{19}$ \GeV) the observed value of the Higgs mass
could only be explained by fine-tuning the radiative corrections against the bare mass at the level of one
part in $10^{32}$.

{\bf II. Dark Matter:} Another major motivation for new physics at the
TeV scale comes from dark matter (DM). The existence of DM is well
established from numerous sources, astrophysical and cosmological, and
for recent reviews see e.g.\ \cite{Bertone:2004pz,Feng:2010gw}.  DM is
the dominant component of matter in the universe, yet it interacts
only very weakly with ordinary matter. TeV-scale DM is motivated by
the ``WIMP miracle" paradigm, which is the observation that a stable
TeV-scale particle with weak interactions has the right annihilation
cross section in the early universe in order to account for all of the
DM today.

\bigskip

In this article, we review the status of searches for new physics at
the end of Run I at the LHC and discuss some of the implications.  As
is well-known by now, no BSM physics has been found so far, and there
is a sense in the community that the LHC results are in tension with
naturalness (see Eqn. \ref{eq:TeVscale}).  This lack of new physics, and the re-introduction of the fine-tuning problem that accompanies it, is commonly referred to as the ``little hierarchy problem". By surveying the LHC
searches for new physics, together with their weaknesses, gaps and
loopholes, we will examine the state of naturalness.  

The remainder of this article is organized as
follows. Sec.~\ref{sec:TheoryOverview} provides an overview of
theoretical ideas for BSM physics and reviews the most prominent
models.  This is followed in Sec.~\ref{sec:StatusOfLimits} with the
current status of searches from the ATLAS and CMS experiments.  In
Sec.~\ref{sec:Implications} we assess the implications for specific
BSM models and review re-interpretations of LHC results in the
literature.  We attempt to enumerate the weaknesses in current
searches and loopholes in their interpretation.  Finally, in
Sec~\ref{sec:Outlook} we conclude with a brief look towards the future
of the LHC at close to its design energy of $\sqrt{s}=14$~TeV.

\section{THEORY OVERVIEW}
\label{sec:TheoryOverview}

\subsection{Solutions to the Hierarchy Problem}

As discussed in Sec.~\ref{sec:Introduction}, a ``natural" SM Higgs
mass implies new physics at the TeV scale.  However, this alone is not
sufficient to completely solve the hierarchy problem.  Generally, the
new physics must also have some additional structure, such as new
symmetries or strong dynamics, that shields the Higgs mass against
quadratic divergences from even higher scales. In this work, we focus
on two well-studied frameworks for solving the hierarchy problem --
supersymmetry (SUSY) and composite Higgs.

\subsubsection{Supersymmetry}
\label{sec:susytheory}

Supersymmetry (SUSY) is the prime example of a solution to the
hierarchy problem based purely on symmetries. (Composite Higgs models, to be reviewed in section \ref{sec:comphiggs}, use a combination of symmetry and strong dynamics.)
The basic idea of SUSY is that it groups bosons and fermions
into ``supermultiplets," such that particles in the same
supermultiplet have the same properties apart from their spin. In this way, the spin 0
Higgs boson is related to a spin 1/2 Higgsino. Since the Higgsino mass
is protected by chiral symmetry in the same way as the SM fermions,
the Higgs mass becomes protected as well. In practice, what happens is
that the quadratically divergent loop corrections to the Higgs mass in
the SM are canceled by corresponding loops of superpartners. For
instance, the one-loop top diagram that leads to Eqn.~(\ref{eq:TeVscale})
is canceled by a one-loop top squark diagram.

In the Minimal Supersymmetric Standard Model (MSSM), the particle
content of the SM is approximately doubled.\footnote{For an excellent
  review of the MSSM and SUSY phenomenology,
  see~\cite{Martin:1997ns}.} For every SM fermion, there is a spin 0
counterpart, e.g. for a quark $q$ there is a squark $\tilde q$ and for
a lepton $\ell$ there is a slepton $\tilde\ell$. For every SM gauge
boson, there is a spin 1/2 counterpart, gluons to gluinos $\tilde g$,
$W$, $Z$ and $\gamma$ to wino, zino and photino,
respectively. Finally, for the SM Higgs, the MSSM enlarges it to two
Higgs doublets $H_u$ and $H_d$, and they each have spin 1/2
superpartners called Higgsinos. Under EWSB the Higgsinos mix with the
wino, zino and photinos; these become charginos $\tilde\chi^\pm$ and
neutralinos $\tilde\chi^0$.  Some cross sections for SUSY particle
production are shown as a function of mass in Fig.~\ref{fig:susyxsec}
for $\sqrt{s}=8$~TeV and $\sqrt{s}=$13-14~TeV.

\begin{figure}
  \begin{center}
\includegraphics[width=0.8\textwidth]{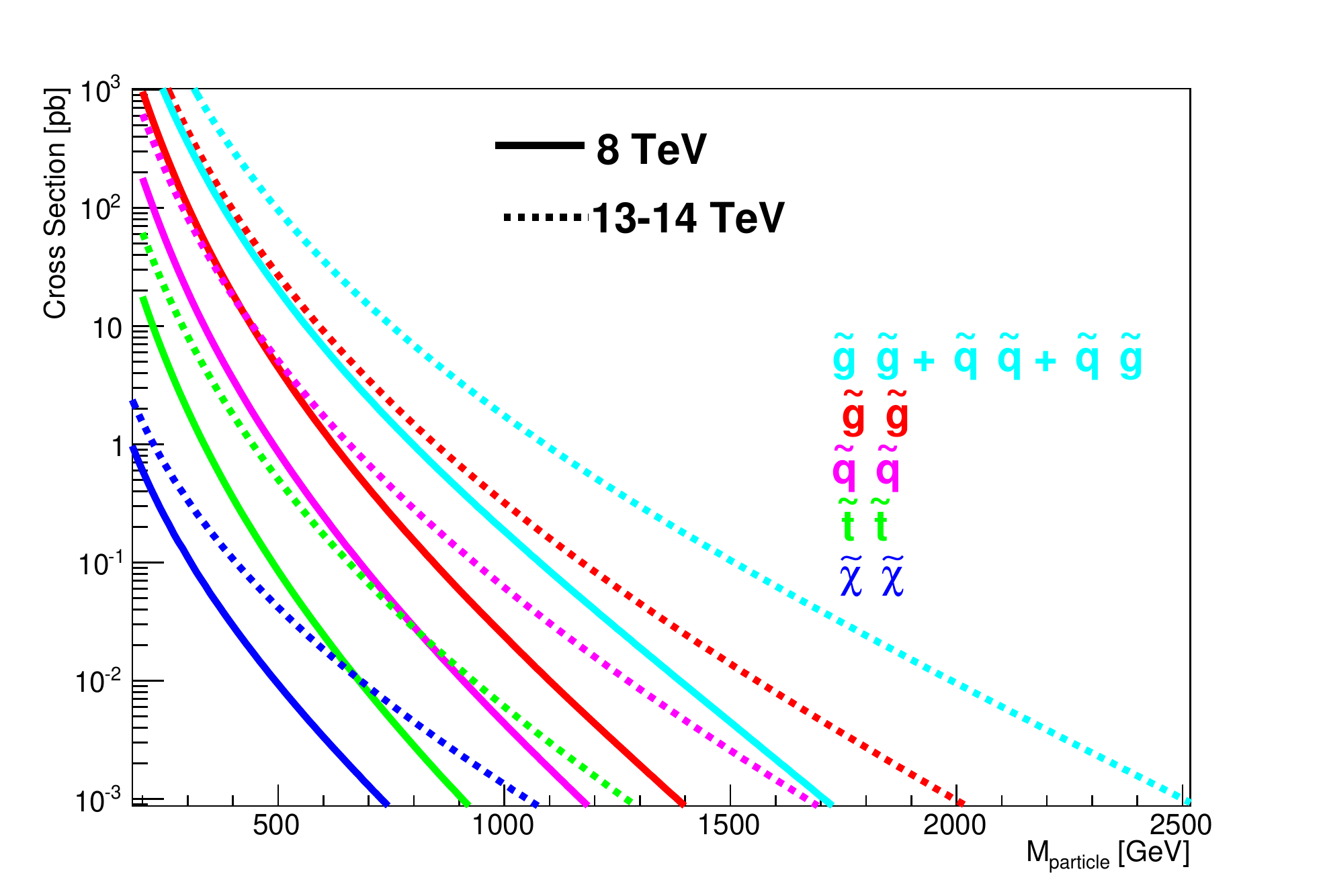}
\caption{Cross sections for SUSY particle production.  at
  $\sqrt{s}=8$~TeV and 13-14~TeV.  The colored particle cross sections are from \textsc{nll-fast}~\cite{Beenakker:2011fu} and evaluated at  $\sqrt{s}=8$~TeV and 13~TeV; the electroweak pure higgsino cross sections are from \textsc{prospino}~\cite{Beenakker:1996ed} and evaluated at $\sqrt{s}=8$~TeV and 14~TeV.  The electroweak pair production cross section is sensitive to mixing, and the higgsino cross sections (shown in the figure) are approximately a factor of 2 lower than the pure wino case. } 
\label{fig:susyxsec}
\end{center}
\end{figure}

One general problem for the MSSM and SUSY is the proliferation of
parameters. 
The MSSM certainly includes the usual Yukawa couplings ${\bf Y_{u,d,\ell}}$ of
the SM:
\begin{equation}
W_{Yukawa} = H_u Q {\bf Y_u} U+H_d Q {\bf Y_d} D + H_d L {\bf Y_\ell} E.
\end{equation}
Here $Q$, $U$, $D$, $L$ and $E$ denote ``superfields" containing the SM fermions and their superpartners. However, even in the supersymmetric limit, there are many more
possible renormalizable interaction terms:
\begin{equation}
W_{RPV} =\lambda_{ijk} L_iL_jE_k +\lambda'_{ijk} L_iQ_jD_k+ \lambda''_{ijk} U_iD_jD_k + \mu'_i L_i H_u .
\end{equation}
Here $i,j,k$ are flavor indices that run over the three generations of the SM. These new interactions all involve at least one superpartner, so they
are not present in the SM. They generically violate baryon and lepton number, as well as the approximate flavor and CP symmetries of the SM. Thus there are stringent constraints on these couplings, see e.g.\ \cite{Barbier:2004ez} for an overview.
These dangerous interactions
can all be forbidden by imposing an additional discrete ${\Bbb Z}_2$ symmetry on the MSSM called R-parity, which assigns charge 1 to
all superpartners and charge 0 to all ordinary SM particles.

One crucial byproduct of R-parity is that the lightest superpartner
(LSP) is absolutely stable. Cosmological bounds strongly suggest that
the LSP must be neutral, in which case it could be a WIMP DM
candidate. At the LHC, pair-produced superpartners\footnote{Single production is forbidden by
  R-parity.}  cascade decay down to the LSP, producing jets and
leptons along the way. The LSP escapes the detector like a heavy
neutrino, implying a missing transverse energy (\met)
signature. Therefore, searches for jets and/or leptons and \met\ are
among the best ways to search for SUSY at the LHC.

Of course, it is possible to reduce or even eliminate the
\met\ signature by compressing the LSP mass against other particles, or
elongating the decay chain, or turning on a small amount of R-parity violation (RPV). As the \met-based searches set stringent limits, there is increasing
interest in such scenarios.

In addition to the RPV supersymmetric operators, there are also many
dangerous R-parity-conserving interactions when SUSY breaking is taken
into account. If SUSY were unbroken, the superpartners would be
degenerate with their SM counterparts. Since the superpartners have
not yet been observed, SUSY must be realized as a spontaneously broken
symmetry in Nature. The superpartner spectrum is parametrized by
the soft SUSY-breaking Lagrangian:
\begin{eqnarray}
{\mathcal L}_{soft} && =\sum_{\tilde f=\tilde q,\tilde u,\tilde d,\tilde l,\tilde e} \tilde f^\dagger {\bf m_{\tilde f}^2} \tilde f + \left(\sum_{r=1}^3 M_r \lambda_r\lambda_r+c.c.\right)\\ \nonumber
&& +\left(m_{H_u}^2 |H_u|^2 + m_{H_d}^2|H_d|^2+ B\mu H_u H_d + c.c.\right)
\\  && + \left(H_u \tilde q A_u \tilde u + H_d \tilde q A_d \tilde d + H_d \tilde \ell A_\ell \tilde e +c.c.\right).\nonumber
\end{eqnarray}
The terms on the first line describe the squark, slepton and gaugino
soft masses; the terms on the second line describe the Higgs soft
masses, and the terms on the third line describe the trilinear soft
terms (the ``A-terms").  While the soft masses do preserve baryon and lepton number, they generically violate the approximate flavor and CP symmetries of the SM. And unlike the RPV couplings, there is no symmetry that can forbid such terms. This is known as the SUSY flavor and CP problem.

Gauge Mediated SUSY Breaking (GMSB) is the most promising solution of
the SUSY flavor problem.\footnote{For a review of GMSB and original references,
  see~\cite{Giudice:1998bp}.} It postulates that SUSY breaking is only
communicated to the MSSM via the SM gauge interactions. Since these
are flavor blind, the resulting soft masses will be flavor
blind. Minimal GMSB models were first constructed in the seminal works
of \cite{Dine:1993yw,Dine:1994vc,Dine:1995ag}. The most general
parameterization of GMSB was formulated in
\cite{Meade:2008wd,Buican:2008ws}. One distinctive feature of GMSB is
that the gravitino is the LSP and is typically quite light,
$m_{gravitino}\lesssim{\rm keV}$. The lightest MSSM superpartner
(which in GMSB can be any sparticle) becomes the next-to-lightest
superpartner (NLSP) and it decays to  its SM
counterpart and the gravitino, e.g. $\tilde B \to \gamma +\tilde G$. So there are
spectacular signatures such as $\gamma\gamma$ + \met,
$\gamma+\ell$+\met, and multileptons+\met. The NLSP lifetime is also a
free parameter in principle, and the lifetime can range from prompt,
to displaced, to detector-stable. In the latter case there are many
powerful and inclusive searches for CHAMPS and R-hadrons, while for
displaced signatures currently there are very few so far (see
Sec.~\ref{sec:displaced}).

 The complexity of SUSY parameter space is problematic for both
 theorists and for experimentalists who must design searches and set
 limits on specific slices of this parameter space. As it is
 impossible to cover the entire parameter space by simulation, several
 complementary approaches are taken when estimating the sensitivity of
 the searches to SUSY signals.  In the first approach, complete SUSY
 models are simulated; these models typically impose boundary
 conditions at a high energy scale, reducing the number of parameters
 to about five, and making it realistic to scan the parameter space by
 brute force. Examples are
 MSUGRA/CMSSM~\cite{Chamseddine:1982jx,Barbieri:1982eh,Ibanez:1982ee,Hall:1983iz,Ohta:1982wn,Kane:1993td},
 and minimal GMSB and anomaly-mediated SUSY
 (AMSB)~\cite{Giudice:1998xp,Randall:1998uk} models.

The second approach is referred to as ``simplified models''
\cite{Alwall:2008ve,Alwall:2008va,Alwall:2008ag,Alves:2011wf}, and is commonly used in BSM
searches in general.  As applied to SUSY, the decay cascades are
simplified by setting the masses of most SUSY particles to multi-TeV
values, putting them out of range of the LHC.  The decay cascades of
the remaining particles to the LSP, typically with zero or one
intermediate step, are characterized only by the masses of the
participating particles, allowing studies of the search sensitivity to
the SUSY masses and decay kinematics.  A typical limitation of this
approach is the fact that all events decay through only one chain;
even within one decay chain, once the number of states exceeds two,
various assumptions are typically imposed on the relationship between
the masses, both to save computing time and to simplify the
visualization of the final results.

One popular example of a simplified-model-type scenario are ``natural"
or ``effective" SUSY models.\footnote{For recent reviews on ``natural
  SUSY'' models and original references, see
  e.g.\ \cite{Feng:2013pwa,Craig:2013cxa}.} It was noticed long ago
that not all superpartners are equally important for the fine-tuning
problem of the electroweak (EW) scale. For instance, in the MSSM,
at tree-level only Higgsinos contribute to fine-tuning, at one-loop stops are most important, and at two-loops gluinos are most important.  The idea of
``natural SUSY" is to imagine that all but Higgsino, stop and gluino
are decoupled from the LHC (in practice, heavier than $\sim 10$~TeV). This would be the minimal superpartner
spectrum necessary to postpone the hierarchy problem to much higher
scales.  Aside from having a much simpler parameter space, the main
benefit of ``natural SUSY'' is significantly weakened LHC constraints,
primarily because decoupling the first generation squarks greatly
decreases the SUSY production cross section.

\subsubsection{Composite Higgs}
\label{sec:comphiggs}

Composite Higgs models attempt to solve the hierarchy problem through a combination of strong dynamics and symmetry.\footnote{For a recent review of composite Higgs models and
  many original references, see~\cite{Bellazzini:2014yua}.}   First, by
positing that the Higgs is a composite bound state of an additional strongly-coupled
sector (and not an elementary
scalar), these models cut off the quadratic
divergence of the Higgs mass at the compositeness scale. Direct searches, precision EW tests, and
flavor tests all constrain the compositeness scale to at least the
multi-TeV range. Therefore, modern composite Higgs models (especially
after the Higgs discovery) generally also equip the composite sector with an approximate global symmetry $G$, in order to explain why the Higgs is a narrow, light state apparently well-separated from the
other resonances of the composite sector. When $G$ is spontaneously
broken to a subgroup $H$ at some scale $f\sim 1$~TeV, the SM Higgs
emerges as a pseudo-Nambu-Goldstone boson (PNGB), much like the pion
of QCD.  In order to better satisfy precision EW constraints, it is
generally assumed that $H$ contains the custodial $SO(4)$
symmetry. $G$ is also explicitly broken by the gauging of $SU(2)\times
U(1)$ and by the SM Yukawa couplings. These radiatively generate a
potential for the Higgs.

One popular and well-studied example is the ``Minimal Composite Higgs
Model" (MCHM) \cite{Agashe:2004rs}. Here the strong sector has a
global $G=SO(5)$ symmetry that is spontaneously broken down to
$H=SO(4)$. After gauging a $SU(2)\times U(1)$ subgroup of $SO(5)$, there is one light PNGB
with the correct quantum numbers to be a SM-like Higgs.

As in QCD, composite Higgs models predict a tower of resonances
starting somewhere around the compositeness scale. Since the composite
sector transforms under $SU(2)\times U(1)$, some of these resonances
will carry EW quantum numbers.\footnote{One can also consider the
  phenomenology of spin 1 colored resonances. These are not required
  in composite Higgs models, but generically arise as KK gluons in
  extra dimensional models.} The lowest EW spin 1 resonance is usually
called the $\rho$ by analogy with QCD. The strongest limits on the
$m_\rho$ are 2-3~TeV from EW precision constraints (for a pedagogical
overview of this, see \cite{Contino:2010rs}). This would put them out
of reach of the LHC. Nevertheless, searches for them are ongoing
(generally phrased as searches for $W'$ particles). These resonances
mix with the $W$ and $Z$ bosons so they can be produced from Drell-Yan
and Vector Boson Fusion (VBF). They can decay in many ways, including
$WW$, $WZ$, $Wh$, $Zh$, $t\bar t$ and $t\bar b$.

In addition to EW vector resonances, there are also generally colored
fermionic resonances in composite Higgs models. These are necessary to
realize the ``partial compositeness" scenario that alleviates the
flavor problem of composite Higgs models. Here the SM fermions $q$ are
assumed to mix with fermionic resonances ${\mathcal O}_q$ from the
composite sector:
\begin{equation}
{\mathcal L}\supset \lambda_q q {\mathcal O}_q  + \dots
\end{equation}
The 3rd generation is assumed to couple most strongly to the fermionic
resonances; thus, the lightest one of these are usually referred to as
``top partners". 

For the phenomenology of composite Higgs top partners, we refer
to~\cite{DeSimone:2012fs,Aguilar-Saavedra:2013qpa}. The former paper focuses on top partners in the MCHM, while the latter paper takes a more model-independent point of view. Starting from complete multiplets of the global symmetry $G$ of the composite sector, and decomposing them into EW representations, a number of distinct top partners can emerge. This
includes particles (often denoted by $T$ and $B$) with the same quantum numbers as the top and bottom quarks, as well as particles with exotic electric charges such as
5/3 (often denoted as $T_{5/3}$). These can be pair produced via QCD; then their cross sections
depend solely on their mass and are essentially that of a heavy
vector-like quark. One can also have single production via $Wb$, $Wt$,
$Zb$ and $Zt$ fusion from a gluon-quark initial state, since these top
partners mix with the third generation. This is more model dependent,
since it depends on the unknown mixing parameter; more details are given
in~\cite{DeSimone:2012fs,Aguilar-Saavedra:2013qpa}. Some typical cross sections for top partners produced singly and in pairs are
shown in Fig.~\ref{fig:toppartxsec}. The top partners primarily decay into $Zt$,
$Zb$, $Wt$, $Wb$. Decays involving Higgses are also possible.  So
events have many $b$'s, leptons, jets and \met.

\begin{figure}
  \begin{center}
\includegraphics[width=0.8\textwidth]{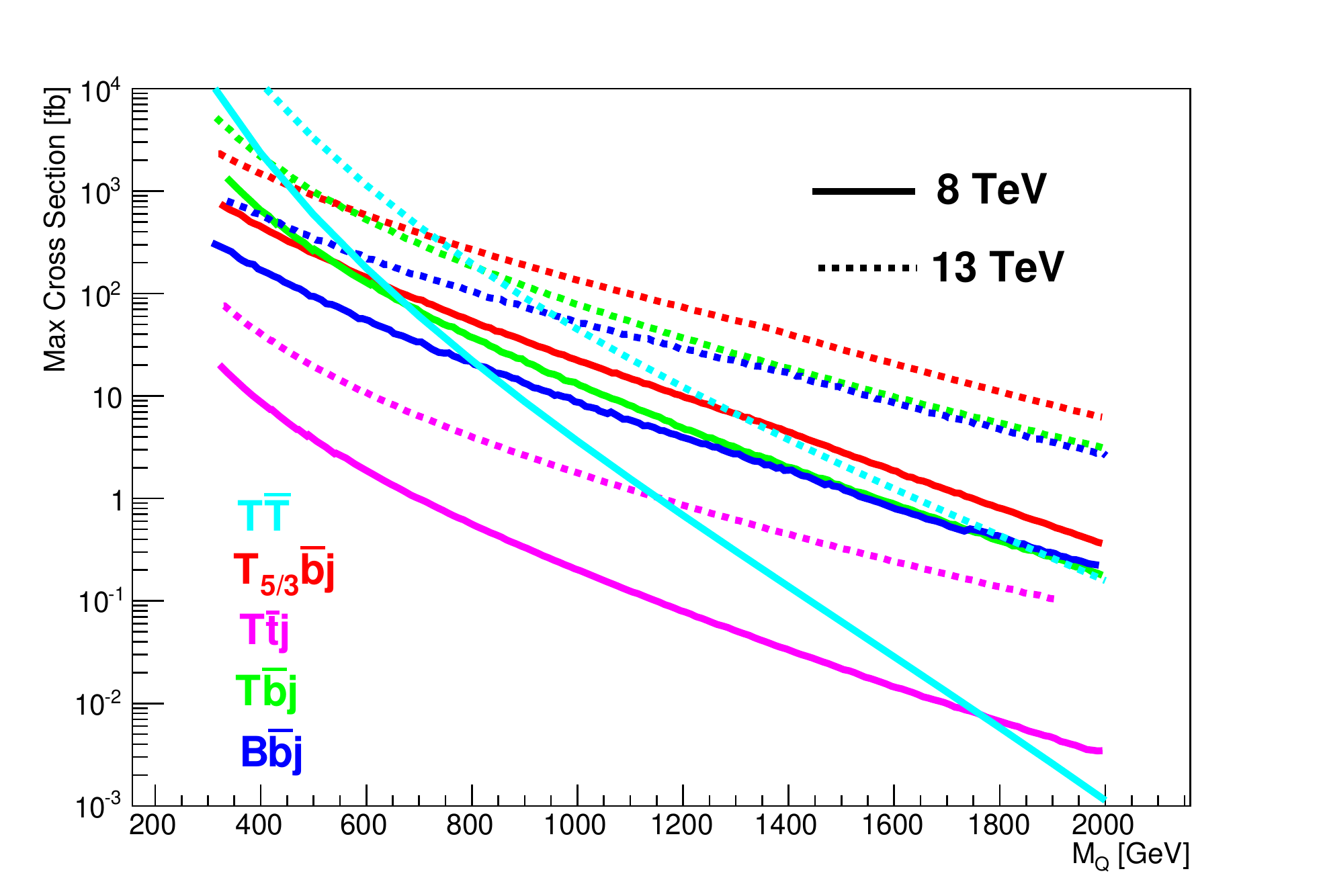}
\caption{Cross sections for single~\cite{Aguilar-Saavedra:2013qpa} and pair~\cite{Aliev:2010zk} top partner production at $\sqrt{s}=8$~TeV and 13~TeV.}
\label{fig:toppartxsec}
\end{center}
\end{figure}

There is an important correspondence  \cite{Verlinde:1999fy,Gubser:1999vj,Verlinde:2000px,ArkaniHamed:2000ds,Rattazzi:2000hs}
between 4D composite Higgs models
and extra-dimensional models such as the Randall-Sundrum model
\cite{Randall:1999ee}. The holographic principle, most notably the
AdS/CFT duality \cite{Maldacena:1997re,Gubser:1998bc,Witten:1998qj}, is
believed to relate weakly-coupled gravitational theories with
strongly-coupled field theories living on their boundaries. 
This has given rise to the hope that composite Higgs models could be rendered calculable through holography. In the simplest 5D constructions, there is an ``IR brane" where the SM
fields are localized, and there is a ``UV brane" where the composite
sector is localized. These branes are separated by a compact
extra dimension; if the spacetime metric in the 5th dimension is
exponentially warped as in RS, then the warping connects the Planck
scale on the UV brane with the TeV scale on the IR brane.
Such
``holographic" composite Higgs models can be weakly coupled in 5D and can describe many of the properties of
a strongly-coupled field theory in a calculable way.  For instance,
the dimensional reduction of a 5D theory down to a 4D theory results
in towers of heavier states, the so-called Kaluza-Klein (KK)
states. These are in direct correspondence with the towers of
resonances expected in a QCD-like theory in 4D.

 In the
original RS construction,
the Higgs is also localized on the IR brane. Therefore it is naturally
of the same size as the KK modes, which in practice must be at least
$\sim 10$~TeV due to flavor and precision EW constraints. Such models
therefore suffer from a severe little hierarchy problem. This
motivated the inclusion of a global symmetry spontaneously broken down
to custodial symmetry, in which the Higgs is a PNGB, both alleviating
the EW precision constraints, and explaining why the Higgs is lighter
than the KK modes \cite{Contino:2003ve,Agashe:2004rs}. In the
extra-dimensional context, the global symmetry on the IR brane is
gauged, and the Higgs propagates in the bulk as the fifth component of
the gauge field.

\subsection{Dark Matter}
\label{sec:darkmattertheory}

As discussed in Sec.~\ref{sec:Introduction}, the connection of dark
matter to the TeV scale (and hence the LHC) proceeds through the
so-called ``WIMP miracle".  If DM is a cold thermal relic, then its
present day annihilation cross section is $\langle \sigma v\rangle
\sim 3\times 10^{-26}$~cm$^3$ s$^{-1}$. Meanwhile, if DM interacts
with the SM through the weak interactions (including Higgs exchange),
then $\sigma \sim {g^4\over m_\chi^2}$. This reproduces the desired
cross section for $m_\chi\sim 1$~TeV.

Despite the connection to the TeV scale, direct searches for DM at the
LHC are inherently challenging. Since DM should be neutral and
colorless, it will show up as missing energy at the LHC. Searches for
direct production must rely on hard initial state radiation (ISR) for triggering (leading to, for example,
monojet+\met and monophoton+\met signatures), which reduces the
rate. 
  Furthermore, if DM is a WIMP, then its cross section is EW and
not strong, further degrading the mass reach.

One popular approach for expressing the results of dark matter
searches in a model-independent way is via effective field theory
(EFT), see e.g.\ \cite{Beltran:2010ww,Rajaraman:2011wf,Fox:2011pm}. By
expressing the dark matter interactions with the SM (e.g.\ quarks) via
effective operators such as ${1\over\Lambda^2}(\bar\chi\chi)(\bar q
q)$, LHC searches can place bounds directly on the mediator scale
$\Lambda$, and they can map these bounds into the same parameter space
as the direct-detection bounds, since they proceed through the same
effective operators. However, it is important to keep in mind that the
EFT approach is only valid for mediator scales $\Lambda\gg m_\chi$, in
practice several TeV for $m_\chi\sim {\mathcal O}(100\,\,{\rm GeV})$
\cite{Fox:2011pm,Shoemaker:2011vi,Busoni:2013lha,Buchmueller:2013dya}. For lighter mediators (e.g.\ the $h$ and $Z$, relevant to actual standard WIMPs), resonant
effects and the general break down of the EFT imply that the limits
set using the EFT can be wildly off compared to those in a genuine UV
completion.

\section{SEARCHES FOR BSM PHYSICS}
\label{sec:StatusOfLimits}

The ATLAS and CMS experiments have extensively searched for BSM
physics using the 7~TeV and 8~TeV datasets in Run I.  So far, these
searches show no evidence for new physics and have set stringent
limits on many BSM models.  The searches are based on distinct
experimental signatures, and while a particular search may report
limits on a particular model, the search is typically sensitive to a
wide range of models.  Both ATLAS and CMS attempt to provide enough
information in their publications so that interested readers can
reinterpret (``recast'') the results in the context of some other
model.

In this section, we briefly summarize the current status of such
searches, focusing on those models most closely tied to naturalness:
Supersymmetry and composite Higgs models (and the related searches for
extra dimensions and black holes (BHs)).  In addition, we present a brief
summary on the searches for DM.

The review is limited to search results at 8 \TeV\ which have been
published or submitted for publication as of the end of February 2014.  Both ATLAS and CMS have
published an extensive array of search results based on 7 \TeV\ data;
most of these results have been superseded by preliminary results from
8 \TeV\ which have been presented at conferences, but which are not
reviewed in this article. The full list of results, both preliminary
and published can be found at \cite{ATLASPublic,CMSPublic}.

\subsection{Supersymmetry}

\subsubsection{Searches for gluinos and 1st/2nd generation squarks}
\label{sec:sqgl}

For a fixed particle mass, gluinos and first-generation squarks
$\tilde q=(\tilde{u},\tilde{d})$ have the largest SUSY production
cross sections at the LHC (see Fig.~\ref{fig:susyxsec}), proceeding
through $pp \rightarrow \tilde{q}\tilde{q}, \tilde{q}\tilde{q}^{\ast},
\tilde{q}\tilde{g}, \tilde{g}\tilde{g}$. They are thus prime
candidates for the most inclusive searches for SUSY.  Squarks
of the first and second generation 
are often assumed to be mass degenerate in LHC
searches (to
better comply with flavor constraints).

In simplified models with very heavy squarks, gluinos decay via
$\tilde{g} \rightarrow q\overline{q}\tilde{\chi}_{i}^{0}$ or
$qq'\tilde{\chi}_{i}^{\pm}$.  If gluinos are very heavy, squarks decay
via $\tilde{q} \rightarrow q\tilde{\chi}_{i}^{0}$ or
$q'\tilde{\chi}_{1}^{\pm}$.  Ignoring additional jets from initial- or
final-state radiation, event topologies with two, three and four jets
are therefore expected for $\tilde{q}\tilde{q}, \tilde{q}\tilde{g}$,
and $\tilde{g}\tilde{g}$ production, respectively.  More complicated
decay cascades lead to larger numbers of jets in the final state.
When electroweak partners are produced in the decay chain, leptons can be present
via the decays $\tilde{\chi}_{1}^{\pm} \rightarrow W^{(\ast)\pm}
\tilde{\chi}_{1}^{0}$ or $\tilde{\chi}_{2}^{0} \rightarrow Z^{(\ast)}
\tilde{\chi}_{1}^{0}$.  The most inclusive searches for SUSY are
therefore based on the presence of multiple jets; zero, one or more leptons;
and \met, where the latter arises (in part) from the two LSP's in the
event.  Useful observables include \met~and $H_{\rm{T}}$, defined as
the scalar sum of the transverse momenta of the jets (and sometimes
the leptons\footnote{Unless otherwise specified, the term leptons will
  refer to electrons and muons.})  in the event.  The sum
$H_{\rm{T}}+\met$, sometimes called the effective mass (\meff),
reflects the mass difference between the initially-produced SUSY
particle and the LSP, and is approximately independent of the details
of the intermediate states in the decay cascade.

The most sensitive search as of the end of February 2014~\cite{Chatrchyan:2014lfa}, covering
the full 8~TeV dataset, uses selection criteria explicitly tuned for
jet multiplicities from 3 up to 8 or more; 36 signal regions, binned
in $H_{\rm{T}}$, $H_{\rm{T}}^{miss}$, and the number of jets, are
considered.  Squarks of the first and second generation below 780
\GeV\ are excluded for LSP masses below 200 \GeV\ in a simplified
model of $pp \rightarrow \tilde{q}\tilde{q}$ where $\tilde{q}
\rightarrow q\tilde{\chi}_{i}^{0}$.
The limit reduces to approximately 450 \GeV\ if only one squark
flavor and chirality is accessible.  Gluinos below 1.1~TeV are excluded
for light LSP masses in a simplified model of $pp \rightarrow
\tilde{g}\tilde{g}$ where $\tilde{g} \rightarrow
q\overline{q}\tilde{\chi}_{1}^{0}$.

Limits on gluino pair production in simplified models with charginos
in the decay cascade are obtained in an ATLAS high multiplicity
jets+\met~ search \cite{Aad:2013wta} with no leptons.  In a model
where $pp \rightarrow \tilde{g}\tilde{g}$ and $\tilde{g} \rightarrow
qq'\tilde{\chi}_{i}^{\pm} \rightarrow qq' W^{(\ast)\pm}
\tilde{\chi}_{1}^{0}$, gluinos below approximately 1 \TeV\ are
excluded for LSP masses below 200 \GeV, assuming the chargino mass is halfway
between the LSP and gluino masses.  The CMS
search~\cite{Chatrchyan:2014lfa} excludes gluinos below 1.2 \TeV\ in a
similar scenario, but also allows for the decay mode $\tilde{g} \rightarrow
q\overline{q}\tilde{\chi}_{2}^{0} \rightarrow q\overline{q} Z
\tilde{\chi}_{1}^{0}$ with equal probability. Searches based on jets
plus \met~and additional leptons reach roughly similar conclusions
in such models.  A CMS search \cite{Chatrchyan:2013fea} based on 34
signal regions with same-sign dileptons binned in $H_{\rm{T}}$, \met,
and the number of ($b$-tagged) jets excludes gluinos below about 900
\GeV\ for light LSP's, approximately independent of assumptions about
the chargino mass.

Table~\ref{tab:compressed}  summarizes the approximate mass reach for
a massless $\ninoone$  for selected analyses that are representative
of an abbreviated set of probed decay modes.   The table also shows
the
maximum value of the $\ninoone$ mass above which no exclusion limits
exist, as well as the minimum mass difference between the initially
produced SUSY particle and the LSP ($\Delta M$) below which no
exclusion limits exist. This is
discussed further in Section~\ref{sec:compressed}.

\subsubsection{Searches for gluinos and 3rd generation squarks}

Searches for gluinos and squarks of the third generation are
particularly well-motivated by the ``natural SUSY'' scenario and have
been carried out in many channels, utilizing various combinations of
($b$-tagged) jets, leptons, $H_{\rm{T}}$ and
\met~\cite{Chatrchyan:2013lya,Chatrchyan:2013fea,Chatrchyan:2013iqa,Chatrchyan:2013wxa,Chatrchyan:2014aea,Aad:2013wta}.
Even before the advent of the ``natural SUSY'' paradigm, it had been recognized
that third generation squarks could be considerably lighter than those
of the first- and second generation due to the possibility of
significant left-right mixing and mass splitting, arising from the large Yukawa coupling.

In the scenario of gluino pair production followed by the three-body
decay $\tilde{g} \rightarrow t\overline{t}\tilde{\chi}_{1}^{0}$ via an
intermediate off-shell top squark, the best published limit as of the end of February 2014
\cite{Chatrchyan:2013iqa} excludes gluinos up to a mass of 1260
\GeV\ for LSP masses ranging up to almost 500 GeV.  Additionally, the
sensitivity is extended for smaller gluino-neutralino mass splittings
in the region $m(\gluino) - m(\ninoone) < 2m(t)$ and $m(\gluino) -
m(\ninoone) > m(W) + m(t)$, where the decay becomes four body and
proceeds through an off-shell top quark.

For gluino decays via on-shell top squarks
(followed by $\tilde{t} \rightarrow t \tilde{\chi}_{1}^{0}$),
exclusion limits are presented fixing the gluino mass to 1000
\GeV\ and varying the LSP and top squark masses or fixing the LSP mass
to a value around 50 \GeV\ and varying the top squark and gluino
masses \cite{Chatrchyan:2013iqa}. In the former model, the 1000
\GeV\ gluino is excluded for all kinematically accessible top squark
and LSP masses, provided the LSP mass is below approximately 520 \GeV.
In the latter model, the gluino mass limit degrades for smaller top
squark masses, but still exceeds 1000 \GeV\ for a top squark mass as
low as 200 \GeV.  For the three-body gluino decay $\tilde{g} \rightarrow
b\overline{b}\tilde{\chi}_{1}^{0}$ via an intermediate virtual bottom
squark, gluinos are excluded up to a mass of 1170 \GeV\ roughly
independent of LSP mass up to 500 \GeV\ \cite{Chatrchyan:2013wxa}.
Other decay modes considered include $\tilde{g} \rightarrow
b\tilde{b}_{1}^{\ast} \rightarrow b \overline{t} \tilde{\chi}_{1}^{+}
\rightarrow b\overline{t}W^{+}\tilde{\chi}_{1}^{0}$ and the RPV decay 
$\tilde{g} \rightarrow
tbs$ \cite{Chatrchyan:2013fea}.

To cover the possibility where the gluinos are beyond reach, searches
for direct production of third-generation squarks have also been made
by both CMS and ATLAS.  A CMS search \cite{Chatrchyan:2013xna} for $pp
\rightarrow \tilde{t}\tilde{t}^{\ast}$, excludes top squarks with a
mass below 630 \GeV\ for a light LSP in a model where both top squarks
decay via $\tilde{t} \rightarrow t \tilde{\chi}_{1}^{0}$ to
unpolarized top quarks, degrading to about 600 \GeV\ for an LSP mass
as high as 200 \GeV.  Similar limits are obtained in a model where
both top squarks decay via $\tilde{t} \rightarrow b
\tilde{\chi}_{1}^{\pm} \rightarrow b W^{(\ast)} \tilde{\chi}_{1}^{0}$
although the limits depend on the mass splitting between the chargino
and the LSP, degrading as the chargino and the LSP become mass
degenerate.  The same CMS search also tackles the three-body decay
$\tilde{t}
  \rightarrow b W \tilde{\chi}_{1}^{0}$
 via a virtual top quark, which can be significant if the mass
 difference between the top squark and LSP is sufficiently compressed.
 An interesting case is the top squark decay $\tilde{t} \rightarrow b
 \tilde{\chi}_{1}^{\pm} \rightarrow b W^{(\ast)} \tilde{\chi}_{1}^{0}$
 where a small mass splitting between the chargino and the LSP renders
 the lepton-based searches ineffective; this topology is covered by an
 ATLAS search \cite{Aad:2013ija} for two energetic $b$-tagged jets
 plus \met\, vetoing events with leptons and additional jets. Top
 squarks with mass below 580 \GeV\ are excluded for LSP masses up to
 approximately 200 \GeV, assuming a mass splitting between the
 chargino and LSP of 5 GeV.  For larger splittings, the limits weaken
 due to the increasing likelihood of successful reconstruction of a
 lepton from the chargino decay.  
 
 Allowing for additional jets in the
 event, ATLAS searches for the process $pp \rightarrow
 \tilde{b}\tilde{b}^{\ast}$ with $\tilde{b} \rightarrow b
 \tilde{\chi}_{1}^{0}$.  The study of a topology where the two
 $b$-tagged jets recoil against a high-\pt\ jet (typically from
 initial-state radiation) gives the analysis sensitivity to compressed
 scenarios where the mass difference between the bottom squark and the
 LSP is relatively small.  Bottom squarks with a mass below 620
 \GeV\ are excluded for small LSP masses, degrading only slightly up
 to an LSP mass of about 200 \GeV.

A CMS search \cite{Chatrchyan:2013mya} considers a
``natural SUSY'' model with GMSB in which only the top squarks and higgsinos are
accessible.  The model considers top squark pair production followed
by $\tilde{t}
\rightarrow t \tilde{\chi}_{1}^{0}$, or
$\tilde{t} \rightarrow t \tilde{\chi}_{2}^{0} \rightarrow t Z^{(\ast)}
\tilde{\chi}_{1}^{0}$, or $\tilde{t} \rightarrow b \tilde{\chi}_{1}^{\pm}
\rightarrow b W^{(\ast)} \tilde{\chi}_{1}^{0}$ and where the
$\tilde{\chi}_{1}^{0}$ decays to $h \tilde{G}$.  One Higgs
boson is required to decay in the diphoton channel while the other is
accepted in either the diphoton or $b\overline{b}$ channels.  Top squark masses
below 360 to 410 \GeV\, depending on the higgsino mass, are excluded.

\subsubsection{Searches for charginos and neutralinos}
\label{sec:gaugino}

Direct pair production of gauginos at the LHC is dominated by
$\tilde{\chi}_{1}^{+}\tilde{\chi}_{2}^{0}$ and
$\tilde{\chi}_{1}^{+}\tilde{\chi}_{1}^{-}$ production in most of the
MSSM parameter space.  One search at 8 \TeV\ has been published as of the end of February 2014, searching for associated
chargino-neutralino production in the trilepton plus \met~channel
\cite{Aad:2014nua}.  Assuming equal mass $\tilde{\chi}_{1}^{+}$ and
$\tilde{\chi}_{2}^{0}$, these gauginos are excluded below 345 \GeV\ in
models where $\tilde{\chi}_{1}^{+} \rightarrow W^{(\ast)+}
\tilde{\chi}_{1}^{0}$ and $\tilde{\chi}_{2}^{0} \rightarrow Z^{(\ast)}
\tilde{\chi}_{1}^{0}$ with a very light LSP, as shown in
Table~\ref{tab:compressed}; the limits are approximately unchanged up
to a LSP mass of about 75 \GeV.  The discovery of a relatively light
Higgs boson also opens the possibility of the decay
$\tilde{\chi}_{2}^{0} \rightarrow h \tilde{\chi}_{1}^{0}$ and the
inclusion of tau leptons in the analysis adds sensitivity via the $h
\rightarrow \tau \tau$ channel.  In models where $\tilde{\chi}_{2}^{0}$ decays exclusively
via $h \tilde{\chi}_{1}^{0}$ and the Higgs boson decays according to
SM branching ratios, degenerate $\tilde{\chi}_{1}^{+}$ and
$\tilde{\chi}_{2}^{0}$ below approximately 140-148 \GeV\ are excluded,
for LSP masses below about 20 \GeV. Results are also interpreted in a
phenomenological MSSM (pMSSM \cite{Djouadi:1998di}) framework where the strongly interacting
SUSY particles, the sleptons and the CP-odd Higgs boson are all
assumed to be out of reach; $\tan\beta$ is set to 10, the gaugino mass
parameters $M_{1}$ is set to 50 \GeV\ and limits are explored as a
function of $M_{2}$ and $\mu$.  A pMSSM scenario where the
right-handed sleptons become kinematically accessible is also
considered.

\subsubsection{Searches for long-lived particles}
\label{sec:longlived}

Searches for long-lived particles are theoretically well
motivated. The possibility of a metastable bound state containing a
gluino or a squark, the so-called R-hadron, was raised already in the
earliest papers on MSSM phenomenology \cite{Farrar:1978xj}.  More
generally, there is a large class of models within SUSY that predict
the existence of long-lived particles, such as: GMSB, AMSB, RPV, and
split SUSY~\cite{ArkaniHamed:2004fb,ArkaniHamed:2004yi}. There are
many reasons that such particles can have highly suppressed decay
rates -- their decays could proceed through very high dimension
operators (GMSB and split SUSY), extremely small couplings (RPV), or
extreme kinematic suppression (AMSB). \footnote{For recent reviews of
  metastable massive particles, see
  Ref.~\cite{Fairbairn:2006gg,Raklev:2009mg}.}

In order
to search for massive particles with long lifetimes, the analyses deal
with ``unconventional'' signatures ranging from displaced vertices,
disappearing tracks, slowly moving particles with a long
time-of-flight, or heavy particles with such long lifetimes that they
decay in the calorimeter out-of-time with the LHC bunch crossing. 

The experimental signature for R-hadrons is complicated by the fact
that they could have a significant probability of undergoing hadronic
reactions in the detector material
\cite{Kraan:2004tz,Mackeprang:2006gx,Mackeprang:2009ad}.  Several
searches are therefore designed, utilizing different portions of the
ATLAS and CMS detectors.  A CMS search \cite{Chatrchyan:2013oca} based
on the full datasets from both 7- and 8~TeV excludes gluinos with mass
below 1233 to 1322~GeV, depending on the interaction model and the
fraction of gluinos that form glueballs and therefore remain neutral
and weakly interacting throughout the transit of the detector.  Top
squark R-hadrons are excluded below a mass of 818 \GeV. Scalar taus
are excluded below 500 \GeV\ when directly and indirectly produced in a minimal
GMSB model; directly produced staus are excluded below 339 \GeV.  Limits are also placed
on Drell-Yan production of leptons with non-standard charge.  An ATLAS
search \cite{Aad:2013gva}, also covering the full 7- and 8
\TeV\ datasets, looks for R-hadrons that have stopped in the
calorimeter material and decay during periods in the LHC bunch
structure without $pp$ collisions.  The complementarity with respect
to the searches described above is model dependent.  Limits on the
gluino, top squark and bottom squark masses of 832, 379 and 344 \GeV,
respectively, are obtained for an LSP mass of 100 \GeV.

Although these searches cover a broad range of mechanisms for
R-hadron interactions, there remains the possibility that a R-hadron
remains neutral through the entire detector.  This case would be
covered by
the search for a monojet plus \met~signal, discussed in
Sec. \ref{sec:DM}, although explicit limits have not been published. (See however \cite{Dreiner:2012gx,Dreiner:2012sh} for a recasting of monojet and SUSY searches for compressed gluino-bino and squark-bino simplified models, which would also be relevant to the all-neutral R-hadron scenario.)

Chargino production via $pp \rightarrow \tilde{\chi}_{1}^{\pm}
\tilde{\chi}_{1}^{0}$+jet or $pp \rightarrow \tilde{\chi}_{1}^{+}
\tilde{\chi}_{1}^{-}$+jet in an AMSB-inspired scenario has been explored by
ATLAS  \cite{Aad:2013yna}.  In these
scenarios, the lightest chargino and the LSP are nearly degenerate
such that the chargino decay proceeds via $\tilde{\chi}_{1}^{+}
\rightarrow \pi^{+} \tilde{\chi}_{1}^{0}$ where the pion has a
momentum of the order of 100 MeV.  The search looks for events in
which an isolated,
high-\pt~charged track ``disappears'' in the ATLAS tracking volume, the
low-momentum pion going unobserved.  An additional jet from
initial-state radiation is required to trigger the event.  In the
minimal AMSB scenario, charginos with a mass below 270 \GeV\ are
excluded. More general limits are derived in the plane of the
chargino mass versus the lifetime.

\subsubsection{Searches for R-Parity Violating Supersymmetry}

ATLAS and CMS have also performed searches for RPV SUSY.  As mentioned
in Sec.~\ref{sec:susytheory}, these signatures show up with
little-to-no \met~ in the event.  As of the end of February 2014, the experiments have
published two such searches both using the full 8~TeV dataset.  The
first from CMS~\cite{Chatrchyan:2013xsw} selects events with at least
three isolated leptons and $b$-jets, and targets signatures which
arise from stop pair production with RPV decays.  Several RPV
couplings are explored, and excludes stop masses for two scenarios
below 1020~GeV and 820~GeV for an LSP mass of 200~GeV.  The second
analysis also from CMS~\cite{Chatrchyan:2013gia} searches for the pair
production of three-jet hadronic resonances, and targets final states
with only-light flavor jets and both light- and heavy-flavor jets.
The results are interpreted in the context of pair produced gluinos
decaying via RPV under two different scenarios, with and without
heavy-flavor.  For the light-flavor decay scenario, gluino masses are
excluded below 650~GeV; for the heavy-flavor decay scenario, gluino
masses are excluded between 200-835~GeV, for the first time.

\begin{table}\noindent\begin{minipage}{\textwidth}%
\def~{\hphantom{0}}
\begin{center}
\caption{Summary of the approximate mass reach for a number of simplified
supersymmetry production/decay channels, for compressed and noncompressed decay
topologies, from the LHC-8 TeV data using up to $\approx 20$ \ifb.}\label{tab:compressed}
\vspace{0.2in}
\begin{tabularx}{\linewidth}{lcccc}%
  \hline
          & min($\Delta M$)\footnote{The minimum mass difference is
    shown, below which mass limits are not evaluated}
  & max $\ninoone$
  mass\footnote{Maximum value of $\ninoone$ mass for which a limit is
    quoted}
  & Mass limit (GeV)
  & Reference \\
Mode  & (GeV)                 & (GeV)                            &  massless $\ninoone$ &\\  
\hline\hline
$\squark\squark \rightarrow q\ninoone~ q\ninoone$ & 175 & 300 & 750 & \cite{Chatrchyan:2013lya}\\
$\gluino\gluino \rightarrow qq \ninoone~ qq \ninoone$ & 25 & 530  &
1160 & \cite{Chatrchyan:2014lfa} \\
\\
$\gluino\gluino \rightarrow tt\ninoone~ tt\ninoone$  & 225 & 580 &
1260 & \cite{Chatrchyan:2013iqa} \\
\hspace{1.0cm}$(m(\stopone) >>m(\gluino))$ & & & &\\
$\gluino\gluino \rightarrow bb\ninoone~ bb\ninoone$  & 50 & 650 & 1170
& \cite{Chatrchyan:2013wxa} \\
\hspace{1.0cm}$(m(\sbottomone) >> m(\gluino))$ & & & &\\
  \\
$\stopone\stopone \rightarrow t\ninoone~ t\ninoone$ & 100 &
  230\footnote{165 GeV for off-shell top} & 630 & \cite{Chatrchyan:2013xna}\\
$\sbottomone\sbottomone \rightarrow b\ninoone b\ninoone$ & 15 & 260 &
  620 & \cite{Aad:2013ija} \\
  \\
$\chinoonep\ninotwo \rightarrow W^{(\ast)}\ninoone Z^{(\ast)}\ninoone$& 25 &
  120 & 345 & \cite{Aad:2014nua} \\
\hline
\end{tabularx}
\end{center}
\end{minipage}
\end{table}

\subsection{Composite Higgs and Extra Dimensions}

As discussed in Section~\ref{sec:TheoryOverview}, composite Higgs
models predict the existence of light ($\lesssim$~1~TeV) colored,
fermionic ``top partners" which couple strongly to 3rd generation
quarks, resulting in signatures with many $b$ quarks, leptons, jets
and \met. In addition, composite Higgs models predict the existence
of EW spin 1 resonances which mix with the $W$ and $Z$ and thus have
phenomenology similar to $W'$s and $Z'$s. These are generally
constrained to lie above 2-3~TeV by precision EW tests. Finally,
colored spin 1 resonances (the so-called KK gluons) are generically
present in many concrete (holographic) models of composite Higgs, and
these are much less constrained by precision EW. Searches for all of
these resonances are ongoing at the LHC.

So far the experiments have performed targeted searches for: top
partners with charge $5/3$ ($T_{5/3}$) decaying to a top quark and a
$W$ boson~\cite{Chatrchyan:2013wfa}; top partners with charge $2/3$
($T$) decaying to either a $b$ quark and a $W$, a top quark and
a $Z$ boson, or a top quark and a Higgs
boson~\cite{Chatrchyan:2013uxa}; a $W^{\prime}$ decaying to $t\bar
b$~\cite{Chatrchyan:2014koa}; excited tops decaying to a top quark and
a gluon~\cite{Chatrchyan:2013oba}; and generic searches for anomalous
production of $t\bar{t}$ that constrain $Z'$s and KK
gluons~\cite{Chatrchyan:2013lca}.

In each of these searches $t \bar{t}$ is a large background and they
therefore exploit the kinematic differences between the particles and
the top background.  A variety of distinguishing kinematic variables
are used in these analyses, and examples include: invariant mass
reconstruction (such as $M_{tb}$ in the $W^{\prime} \rightarrow t\bar
b$ search, or $M_{t\bar{t}}$ in the anomalous $t\bar{t}$ production
search), $H_T$, and more sophisticated multivariate tools (such as a
boosted decision tree in the $T$ search).

No evidence for top partners or vector resonances has been found in
the LHC data so far, and stringent limits have been placed on a
variety of scenarios.  The exclusion limits on heavy resonances such
as $W^{\prime}$, $Z^{\prime}$ and Randall-Sundrum Kaluza-Klein (RS KK)
gluons are in the 2~TeV range, whereas limits on vector-like top
partners are in the 600-800 range.  In Table~\ref{tab:b2gbh} we
summarize the current status of the 8~TeV searches from the LHC
experiments.

In models of extra dimensions, access to quantum gravity at the TeV
scale can lead to large rates for black hole (BH) production at the LHC
\cite{Banks:1999gd,Giddings:2001bu,Dimopoulos:2001hw}. Searches at the LHC have
been performed for both ``small" quantum BHs and
``large" semiclassical BHs.  In the latter case, the
experimental signature would present itself as a high multiplicity of
high $p_T$ particles, produced as the semiclassical BH evaporates via
Hawking radiation.  Meanwhile, ``small" quantum BHs would only have
enough energy to decay to small multiplicity of particles, yielding a
signature which is distinct from the semiclassical case.

The ATLAS and CMS experiments have searched for both such signatures
for BHs and the current status is summarized in Table~\ref{tab:b2gbh}.
In the CMS search~\cite{Chatrchyan:2013xva}, 12 fb$^{-1}$ of the 8~TeV
data is used and explores events with large particle multiplicities.
Model-dependent limits are obtained for several scenarios (such as the ADD model and string balls) and assumptions (such as
rotating or non-rotating BH, or the number of extra dimensions) and
are all in the roughly multi-TeV range.  In addition,
model-independent limits are provided as a function of the scalar sum
of the $p_T$s of all the final-state objects, and the particle
multiplicity, which allows straight-forward possible re-interpretation
of the results.  ATLAS has performed three searches for extra
dimensions: two searches target quantum BH in either the lepton+jets
final state~\cite{Aad:2013gma} or the photon+jet final
state~\cite{Aad:2013cva}; and the third search targets an ADD model
with like-sign dimuons~\cite{Aad:2013lna}.  Again, the lower mass
limits are reported for several different scenarios and model
assumptions and are in the 5~TeV range, as summarized in
Table~\ref{tab:b2gbh}.

\begin{table}
\begin{center}
\caption{Lower mass limits (or ranges of lower limits) 
on top partners and vector resonances from
  Composite Higgs models and limits on extra dimensions.  These LHC
  searches use the 8~TeV data up to $\sim$20~fb$^{-1}$, unless
  otherwise noted. The limit is on the mass of the heavy particle
  searched for, unless otherwise indicated.  The abbreviations are
  defined in the text.}
\vspace{0.2in}
\begin{tabular}{llll} \hline
Search & Signature &  Limit &  [Ref.] \\ \hline
$W^{\prime} \rightarrow t b$ resonances & $\ell$+jets & 2.05 TeV & \cite{Chatrchyan:2014koa} \\ 
$T_{5/3} \rightarrow t W$ & same-sign dileptons & 800 GeV & \cite{Chatrchyan:2013wfa} \\ 
$T_{2/3} \rightarrow b W, tZ, tH$ & $\ell$+jets and dileptons & 687 - 782 GeV & \cite{Chatrchyan:2013uxa}  \\ 
$t^* \rightarrow t g$ & $\ell$+jets & 803 GeV & \cite{Chatrchyan:2013oba} \\ 
Anomalous $t\bar{t}$ production & $\ell$+jets and all-hadronic & ${Z^{\prime}}:$ 2.1 - 2.7 TeV & \cite{Chatrchyan:2013lca}\\ 
 & & RS KK gluon: 2.5 TeV & \\ \hline
microscopic BH & large $S_T$, jets, $\ell$s, $\gamma$s and \met & 4.3-6.2~TeV &  \cite{Chatrchyan:2013xva} (12~fb$^{-1}$) \\
quantum BH & $\ell$+jets & 5.3~TeV &  \cite{Aad:2013gma} \\
quantum BH & $\gamma$+jets & 4.6~TeV &  \cite{Aad:2013cva} \\
ADD BH & like-sign dimuons & 5.1-5.7~TeV & \cite{Aad:2013lna} \\
\hline
\end{tabular}
\label{tab:b2gbh}
\end{center}
\end{table}

\subsection{Direct Searches for Dark Matter}
\label{sec:DM}

In addition to the searches for SUSY described above which provide a
possible candidate for dark matter (e.g.\ the LSP), the LHC
experiments have searched for direct pair production of WIMP DM
particles. \footnote{The pair production is a consequence of the
  assumed parity symmetry that keeps the WIMPs stable.} Since the
WIMPs escape the detector without interacting, in order to be able to
trigger on them, they are also assumed to be produced in association
with a particle $X$, where $X$ can be a jet, photon, or a vector
boson.  This gives rise to a mono-$X$+\met\ signature.  Although
searches for DM have been extensively explored with the 7~TeV data and
in preliminary results from 8~TeV data, there is currently only one
such published result as of the end of February 2014.  The recent publication from
ATLAS~\cite{Aad:2013oja} uses the full 8~TeV dataset and searches for
a single hadronic large-radius jet whose mass is consistent with that
of a $W$ or $Z$ boson, and is recoiling against large missing
transverse momentum. Limits on the DM-nucleon scattering cross section
as a function of the mass of the DM particle are set in the context of
EFT, as described in Section~\ref{sec:darkmattertheory}, and are
reported for both spin independent (vector-like) and spin-dependent
(axial-vector-like) scattering.  The results are also presented
alongside those from direct dark matter detection experiments,
however, it is important to note that direct comparisons using the EFT
approach has important caveats as was briefly reviewed earlier in
Section~\ref{sec:darkmattertheory}.

\section{IMPLICATIONS}
\label{sec:Implications}

\subsection{Implications for Specific Models}

Prior to the LHC turn-on, a number of benchmark ``complete" SUSY
models were studied both in the theoretical literature and by the LHC
experiments.  In this section we review the current status of some of these
models in the aftermath of Run I.

Perhaps the most studied benchmark model has been the constrained MSSM
(CMSSM). Here the vast MSSM parameter space is reduced to just five
parameters set at the GUT scale: $m_0$ (common sfermion mass),
$m_{1/2}$ (common gaugino mass), $A_0$ (common trilinear soft term),
$\tan\beta$, and the sign of $\mu$.  Fits to existing data in the
CMSSM framework hinted at a SUSY mass scale in the hundreds of
\GeV\ range, although this value was primarily driven by the deviation
from SM expectations in the measurement of the anomalous magnetic
moment of the muon (see e.g.\ the discussion in
\cite{Trotta:2008bp,Buchmueller:2009fn}).  At the end of Run I, the
situation of the CMSSM is much different. For the most up-to-date
fits and references to earlier work, see
\cite{Bechtle:2013mda,Buchmueller:2013rsa}. The Higgs at 125~GeV (to
a large extent) and the direct LHC searches (to a lesser extent) have
pushed the superpartners up much higher in mass, with stops $\gtrsim
750-1000$~GeV and gluinos and squarks $\gtrsim 1500-2000$~GeV now
being required.

Another popular ``complete" model with a greatly reduced parameter
space is Minimal Gauge Mediation (MGM) \cite{Dine:1993yw,Dine:1994vc,Dine:1995ag}. Here some
number of ${\bf 5}\oplus{\bf \bar 5}$ messengers generate
flavor-diagonal soft masses at a messenger scale $M$. The soft masses
are proportional to a common scale $\Lambda$. Together with the
gravitino mass $m_{3/2}$ and the usual Higgs parameters $\tan\beta$
and ${\rm sign}(\mu)$, these form the parameter space of MGM. Here the
Higgs mass constraint alone is enough to push the soft masses into the
multi-TeV range and basically out of reach of even the 14~TeV LHC
\cite{Ajaib:2012vc}.

The common theme to both of these examples is that for simple,
constrained models {\it based on the MSSM}, requiring a 125 GeV Higgs
mass is generally more powerful than any individual LHC search in
constraining the parameter space. In the MSSM, the Higgs mass is
bounded at tree-level to be less than $m_Z$. So $m_h=125$~GeV requires
\cite{Hall:2011aa,
  Heinemeyer:2011aa,Arbey:2011ab,Draper:2011aa,Carena:2011aa} large
radiative corrections: either stops $\gtrsim 1$~TeV with large
$A$-terms (the ``max-mixing scenario"
~\cite{Casas:1994us,Carena:1995bx,Haber:1996fp}), or in the absence of
$A$-terms, very heavy stops $\gtrsim 8-10$~TeV. In constrained models
such as CMSSM and MGM, where all of the superpartner masses are
controlled by just a few parameters, requiring such heavy stops
generally also pushes the rest of the spectrum out of reach. 

Of course, even models based on the MSSM need not be so constrained as
the CMSSM and MGM. For example, since the Higgs discovery, much effort
has been devoted to extending GMSB models in order to realize large
$A$-terms
\cite{Evans:2012hg,Kang:2012ra,Craig:2012xp,Abdullah:2012tq,Craig:2013wga,Byakti:2013ti,Evans:2013kxa,Calibbi:2013mka,Jelinski:2013kta,Knapen:2013zla,Kumar:2014uxa}. Such
models, while often still out of reach of 7-8~TeV LHC, could be
accessible at 14~TeV. One can also consider going beyond the MSSM;
adding additional interactions can circumvent the tree-level
bound. For instance, the Higgs mass can be raised by couplings to
additional singlets (as in the NMSSM\footnote{For a review of the NMSSM see
e.g.\ \cite{Ellwanger:2009dp}.}) or by adding additional gauge
interactions \cite{Batra:2003nj}.

\subsection{Broader Implications}
\label{sec:recast}

Given the vast multitude of LHC searches for new physics that have
been performed by ATLAS and CMS in many different channels, each
setting a limit on a specific model or simplified model, it can be
challenging to draw more general lessons. Many theorists have
endeavored to reinterpret or ``recast" LHC analyses, in order to
understand broader implications, or in order to understand the
constraints on their favorite models and potential loopholes in
existing searches. \footnote{For a comprehensive list of references to
  such papers (and an interesting proposal for a potentially efficient
  shortcut to recasting), see \cite{Papucci:2014rja}.} In this short
review, we cannot do justice to the many recasting works out
there. Instead we will attempt to highlight some recent examples that
make use of 8~TeV LHC results. In this section we concentrate on
reinterpretations that reach broader conclusions, while
Sec.~\ref{sec:loopholes} is devoted to a general discussion of
loopholes.

Perhaps the largest single theme driving recasting research is
``natural SUSY''. As discussed in Sec.~\ref{sec:TheoryOverview}, in
the MSSM, the Higgsino, stop and gluino are most important for
naturalness in SUSY. Much effort has been devoted to recasting LHC
analyses for simplified models motivated by ``natural SUSY'',
beginning with the work of
\cite{Kats:2011it,Essig:2011qg,Kats:2011qh,Brust:2011tb,Papucci:2011wy}. More
recently, for gluinos, an extensive study \cite{Evans:2013jna},
reinterpreting nearly all relevant Run I LHC searches concludes that a
gluino with a mass below approximately 1~TeV is very likely excluded
for a very broad range of scenarios (RPC, RPV, hidden valleys, top
dilution, etc.), provided that the gluino produces either \met\ or top
quarks in the decay chain. If the tops and \met\ can be diluted
sufficiently such that the final state is all hadronic, then the limit
on the gluino could be lowered to $m_{\tilde g}\gtrsim 800$~GeV.

In \cite{Kowalska:2013ica}, a study of ``natural SUSY'' production
recasted three 8~TeV LHC searches and found that $m_{\tilde g}\gtrsim
1.2$~TeV and $m_{\tilde t}\gtrsim 700$~TeV for $m_{\tilde H}\lesssim
300$~GeV for a few simplified models with tops and \met\ in the final
state. See also \cite{Kribs:2013lua,Han:2013kga} which obtained
similar limits on stop production using ATLAS and CMS direct stop
searches.

Other works have focused on the possibility that the LSP decays via
RPV in ``natural SUSY''.  In the comprehensive work of
\cite{Evans:2012bf,Evans:2013uwa}, stop production with RPV was
exhaustively classified. All possible decays, either directly via RPV,
or through an intermediate Higgsino LSP, were classified and an
extensive list of 7 and 8~TeV LHC searches were recasted to derive the
current constraints on all of these scenarios. While in many instances
(e.g.\ those with LLE or LQD decays that give rise to leptons and
\met\ ), the limits on stops were near their kinematic limit $\sim
700$~GeV, many scenarios are also identified (for example $\tilde
t\to\ b\tilde H\to b\tau q q$ via RPV) where there is {\it no current
  limit} on the stop mass from the LHC!

Meanwhile, for composite Higgs models, the focus of recasting has
largely been on fermionic top partners $X$.\footnote{However, see the
  recent work of \cite{Pappadopulo:2014qza}, where some LHC searches
  were recast in terms of limits on composite Higgs vector
  resonances.}  Existing LHC searches use the same-sign dileptons
channel, motivated by decays such as $X\to tW$. A common theme seems
to be that the LHC searches assume 100\% BR's into specific final
states (such as $tW$ or $tZ$), whereas realistic models will have more
complicated, mixed BR's. See
e.g.\ \cite{Berger:2012ec,Cacciapaglia:2012dd,DeSimone:2012fs} for
recasts of 7~TeV same-sign dileptons searches.

For more recent recasting works that use the full 8~TeV dataset, see
e.g.\
\cite{Azatov:2013hya,Reuter:2013iya,Gillioz:2013pba,Matsedonskyi:2014lla}. These
works generalize existing LHC searches to other scenarios. For
instance, \cite{Azatov:2013hya} considers the constraints when both
$T_{5/3}$ and $B$ are light, and can be both singly and pair
produced. It is shown that the reach of existing same-sign dileptons
searches can be increased somewhat from 770 GeV to roughly 850 GeV in
some corners of parameter space. In \cite{Reuter:2013iya}, a
comprehensive survey of the current status of the Littlest Higgs with
T-parity (LHT) model is given. By recasting a number of LHC SUSY
searches and combining with other constraints, a limit of $640$~GeV on
the compositeness scale $f$ is claimed. In \cite{Gillioz:2013pba} a
comprehensive study of bottom-partners in composite Higgs models is
given. Combining current LHC limits with precision constraints, a best
fit point of $v^2/f^2\sim 0.07$ is determined. Finally, in
\cite{Matsedonskyi:2014lla}, the bounds on charge 8/3 top partners are
studied, which could be present even in the MCHM if the top partner is
embedded in a {\bf 14} of $SO(5)$. Recasting same-sign dilepton
searches a limit of $M_{8/3}>940$ GeV is obtained. The CMS BH
search~\cite{Chatrchyan:2013xva} is also considered, which is not yet
competitive.

\subsection{Weaknesses, Loopholes in Current Searches}
\label{sec:loopholes}

Although the searches for new physics beyond the SM performed by the
LHC experiments are broad and comprehensive, there are still some
areas which remain less well explored.  In addition, in some cases,
especially in SUSY searches, there are important assumptions made in
the interpretations of the results which are worth noting.  Here we
attempt to summarize these weaknesses or loopholes, both by the
experiments and in the theoretical literature, and serve as a list of
areas for improvement in future searches.

\subsubsection{Compressed Spectra}
\label{sec:compressed}
In the context of R-parity conserving SUSY models, there is a weakness
in the case of ``compressed SUSY'' where the mass difference between
the initially produced SUSY particle and the LSP is small, leading to
lower \met\ in the final state and lower signal acceptance.  While
compressed spectra may at first seem to require some additional
tuning, examples exist in the literature where this is not the case
\cite{Fan:2011yu,Murayama:2012jh}.

With a few exceptions, ATLAS and CMS do not generally attempt to set
limits in the region of very small mass differences (ranging from 25
to 175 \GeV, depending on the channel) because the signal acceptance
depends strongly on the modeling of initial-state radiation.  Table
\ref{tab:compressed} shows that mass limits are noticeably weaker for
compressed decay spectra. For example, a natural scenario of pair
production top squarks with mass 500 \GeV\ decaying via $\stopone
\rightarrow t\ninoone$ to a LSP of mass 250 \GeV\ is still not ruled
out.  The production of $\chinoonep\ninotwo$ in the ``natural SUSY''
scenario where the electroweak partners are mostly Higgsino-like and
hence highly mass-degenerate, is also unconstrained by the LHC, even
taking into account re-interpretations of LHC monojet searches
\cite{Bharucha:2013epa,Schwaller:2013baa,Baer:2014cua, Han:2014kaa};
current limits on Higgsino LSPs are essentially no better than LEP!
Even with mild compression, searches for electroweak partners all fail
if the LSP has a mass greater than about 100 \GeV.

\subsubsection{Model Assumptions} 

A second loophole in current BSM search limits arises from the
assumptions on decay branching ratios, for example as in the SUSY
simplified models.  Typically the decay is assumed to proceed via a
single channel with 100\% branching ratio.   As of the end of February 2014, the LHC experiments have not presented SUSY search limits as a
function of branching ratio, except in the search for top
squarks~\cite{Chatrchyan:2013xna} where the mass limits are
recalculated under the conservative (but pessimistic) assumption that
either the search in the $\tilde{t} \rightarrow t
\tilde{\chi}_{1}^{0}$ channel or the search in the $\tilde{t}
\rightarrow b \tilde{\chi}_{1}^{\pm}$ dominates. 

As shown in many studies by theorists, mass limits from individual
searches that assume 100\% branching ratios can degrade significantly
when applied to ``realistic" scenarios with complicated branching
ratios.  For example, in \cite{Buchmueller:2013exa}, a study of
``natural SUSY'' spectra recasted four different 7~TeV LHC searches in
exclusive channels (jets+\met\, jets+lepton+\met\, opposite-sign
dileptons+\met\, and same-sign dileptons+\met\ ). Here it is shown that for
a sufficiently complicated spectrum, the limit from any one of these
searches was degraded because of branching ratios, but full
sensitivity could be recovered by combining channels. Similar studies
for direct stop/sbottom production can be found
in~\cite{Barnard:2014joa,Han:2013kga}, where the role of branching
ratios $\tilde{t} \rightarrow t \tilde{\chi}_{1}^{0}$ or $\tilde{t}
\rightarrow b \tilde{\chi}_{1}^{+}$, as well as $\tilde{b} \rightarrow
b\tilde{\chi}_{1}^{0}$ or $\tilde{b} \rightarrow t
\tilde{\chi}_{1}^{-}$, are explored.  Finally,
in~\cite{Howe:2012xe,Arbey:2012fa,Bharucha:2013epa} the effect of the
branching ratios in chargino-neutralino production on current LHC
limits is explored, specifically $\tilde{\chi}_{2}^{0} \rightarrow
\tilde{\chi}_{1}^{0} Z$ vs.\ $\tilde{\chi}_{2}^{0} \rightarrow
\tilde{\chi}_{1}^{0} h$.

Another example which demonstrates the sensitivity of results on the
assumptions of the model is the dependence of SUSY limits on flavor
mixing.  An illustration of this can be found in~\cite{Blanke:2013uia} which studies
the impact of stop-scharm mixing on the top squark bounds, and find
that they can be appreciably lower.  In addition, it is also shown
that large stop-charm mixing can result in interesting and unexplored
signatures such as $t \bar{c}(c\bar{t}) +$ \met, which could be
searched for immediately.

A final example of the dependence on model assumptions we discuss here
is in the SUSY cross section calculation.  For second generation
squarks $\tilde q=(\tilde c,\tilde s)$, essentially only production
via $\tilde{q}\tilde{q}^{\ast}$ is available, since the
$\tilde{q}\tilde{q}$ and $\tilde{q}\tilde{g}$ processes are greatly
suppressed by PDFs unless $q$ corresponds to a valence
quark. Therefore, their rates are generally reduced relative to those
for the first generation squarks \cite{Mahbubani:2012qq}.  The squark
production cross section would also be lowered if the gluino is a
Dirac fermion \cite{Heikinheimo:2011fk, Kribs:2012gx}.\footnote{A
  Dirac gluino also leads to lower fine-tuning.}  Other interesting
examples with lower production cross sections are the ideas of
``folded SUSY" \cite{Burdman:2006tz} and ``twin SUSY"
\cite{Falkowski:2006qq,Chang:2006ra} in which the partners of the
colored SM particles do not carry QCD charges.

\subsubsection{Scans for Gaps in the Coverage} 

Scans of the pMSSM have been used to search for possible gaps in the
coverage of LHC searches for SUSY  \cite{CahillRowley:2012kx,
  CahillRowley:2012rv,CahillRowley:2012cb,
  Conley:2011nn,Sekmen:2011cz,Carena:2012he}. At
the time of Ref. \cite{CahillRowley:2012kx}, corresponding to public
results available in September 2012, the authors conclude that viable
models exist containing first/second generation squarks, gluinos and
third generations squarks with masses below 600, 700 and 400 \GeV,
respectively. The authors of these scans have identified 
mechanisms that allow low-mass pMSSM points to survive
the LHC searches. They essentially confirm loopholes already
described above:  {\it i)} lowering the production cross
section for first/second generation squarks by splitting the mass
degeneracy, {\it ii)} compressed decay chains, and {\it iii)} branches
to multiple final states in the decay cascade.

\subsubsection{Displaced Decays} 
\label{sec:displaced}

As discussed above in section \ref{sec:longlived}, searches for heavy, long-lived particles are well-motivated by a number of theoretical scenarios in SUSY and beyond. The
LHC experiments have searched for such signatures primarily with the 7~TeV data and have only a few results with the 8~TeV data (see
Sec.~\ref{sec:longlived}).  These analyses are typically very
inclusive and powerful and target a large class of models.  However,
these searches cover typical lifetimes $\gtrsim$ a few nanoseconds (10~cm),
whereas intermediate lifetimes of 0.01-1~ns (corresponding to moderately displaced decays in the tracker volume) are much less
constrained.  For example, see~\cite{Meade:2010ji} and \cite{Graham:2012th} for a discussion of the weaknesses and discovery prospects  in the LHC searches for long-lived particles in the context of GMSB and RPV respectively. There are several
challenges to successfully perform these analyses, which include
designing triggers and specialized reconstruction algorithms specific
to the search.  Nevertheless, the LHC experiments would benefit from
designing searches to target this uncovered and more challenging range
of longer-lived particles.

\subsubsection{Searches for Very Unusual Signatures} 

It is crucial that searches for very unusual signatures, in particular
those which may be missed by other searches, be performed by the LHC
experiments.  In addition, a variety of unusual signatures are
predicted by many different models which address naturalness; for
example, see~\cite{Zurek:2010xf} for a recent review.  Although, many
of these ``unexpected'' new signatures have been searched for at 7~TeV, many of them have not yet been repeated by the LHC experiments
with 8~TeV data.  We highlight some examples here.

Hidden Valley models~\cite{Strassler:2006qa} provide solutions to the
hierarchy problem, and postulate a new, low mass hidden sector that is
only very weakly coupled to the SM.  One way this hidden sector could
reveal itself is through the production of many light particles in the
final state and appear as a collimated stream of leptons, which are
referred to as ``lepton-jets''.  (Such signatures were recently
motivated by various astrophysical anomalies in the context of
indirect DM detection \cite{ArkaniHamed:2008qn,ArkaniHamed:2008qp}.)
Searches for lepton-jets have been performed at
7~TeV~\cite{Aad:2012qua, Chatrchyan:2012cg} but have not yet been
repeated or extended in 8~TeV.  Dedicated reconstruction tools are
required for such searches and increasing pile-up can pose an
additional challenge.  

Another class of hidden valley models predict new particles called
``quirks'' which transform under a new strong force called
``infracolor''~\cite{Kang:2008ea}.  These ``quirks'' are quark-like
and fractionally charged.  These were searched for by CMS using the
7~TeV data using the signature of tracks with associated low rate of
energy loss in the silicon tracker~\cite{CMS:2012xi}, but has yet to
be repeated with the 8~TeV data by either experiment.

Despite the fact that such searches for very unusual signatures are
potentially challenging and require dedicated efforts, they should not
be neglected and should continue to be searched for by the LHC
experiments using the 8~TeV data and beyond.

\section{OUTLOOK}
\label{sec:Outlook}

\subsection{Expectations for 14 \TeV}

The LHC is currently shut down (LS1) and data-taking is expected to
resume at close to the design center-of-mass energy of 14~TeV in 2015.  In the
ensuing data-taking period, an integrated luminosity of approximately
100 \ifb\ is expected to be collected by both ATLAS and CMS.  This
will be followed by another shutdown (LS2) during which the injector
chain will be modified to approximately double the instantaneous
luminosity, after which another 200 \ifb\ is expected to be collected.
Currently under discussion are plans to further increase the
luminosity  of the LHC by an additional factor of 2.5, leading to a
total integrated luminosity of 3000 \ifb\ by some time in the 2030's.

Several large-scale planning exercises for the future of high energy
physics have been held in the last two years
\cite{EuropeanStrategyforParticlePhysicsPreparatoryGroup:2013fia,
  Rosner:2014pja,ECFA-13-284}.  The increase in the energy to 14~TeV
extends the kinematic reach to higher mass, while the increase in
integrated luminosity extends the sensitivity for lower mass processes
with small cross section and for difficult kinematic topologies; an
example can be seen in Fig. \ref{fig:susyxsec}.  Both ATLAS and CMS
have made projections of the expected physics performance for
integrated luminosities of 300 and 3000 \ifb.  The ATLAS projections
are mostly derived from fast simulations incorporating
parametrizations of detector performance obtained from detailed
detector simulations, and reoptimizing the analysis. The detailed
detector simulations include the effect of the increased number of
additional inelastic $pp$ collisions (``pile up'') accompanying each
beam crossing in the LHC.  In CMS, simpler extrapolations are made,
based on current physics analyses, and scaling signal and background
by the change in cross sections and integrated luminosities from 8 to
14~TeV.  Table \ref{tab:future} shows the projected mass reach for
some representative processes.

\begin{table}%
\begin{center}
\def~{\hphantom{0}}
\caption{Projections from CMS and ATLAS for the
  5 $\sigma$ discovery reach for selected processes at the 14
  \TeV\ LHC, assuming an integrated luminosity of 300
  \ifb.}\label{tab:future}
\vspace{0.2in}
\begin{tabular}{lccc}%
  \hline
          & Discovery reach   &                  & \\
Mode & 300 \ifb                 &  Comment & [Ref.]\\
\hline\hline
simplified gluino-squark model  &
2700 & $m_{LSP} \approx 0$  &  \cite{ATLAS:2013hta} \\ 
$\gluino\gluino \rightarrow t\overline{t}\ninoone~ t\overline{t}\ninoone$ &
1900 & $m_{LSP} <  \approx 900$ \GeV & \cite{CMS:2013xfa} \\
$\gluino\gluino \rightarrow b\overline{b}\ninoone~
b\overline{b}\ninoone$ &
1900 & $m_{LSP} \approx 0$ &  \cite{CMS:2013xfa} \\
$\stopone\stopone \rightarrow t\ninoone~ t\ninoone$ &
1000 & $m_{LSP} < \approx 200$ \GeV &  \cite{ATLAS-PHYS-PUB-2013-011} \\
$\sbottomone\sbottomone \rightarrow b\ninoone~ b\ninoone$ &
1050 & $m_{LSP} \approx 0$ &  \cite{ATLAS-PHYS-PUB-2013-011}\\
$\chinoonep\ninotwo \rightarrow W\ninoone~ Z\ninoone$ &
500-600 & $m_{LSP} < \approx 100$ \GeV &  \cite{CMS:2013xfa}\\
$\chinoonep\ninotwo \rightarrow W\ninoone~ h\ninoone$ &
400-500 & $m_{LSP} \approx 0$ \GeV &  \cite{CMS:2013xfa}\\
long-lived $\tilde{\tau}_{1}$ & 800 & direct production & \cite{CMS:2013xfa}\\
$TT \rightarrow bW/ tZ/ th (50\%:25\%:25\%)$ &
1000 &  vector-like quarks & \cite{CMS:2013xfa}\\
\hline
\end{tabular}
\end{center}
\end{table}

Several studies have examined the prospects for detecting
mass-degenerate Higgsinos in future LHC runs, based on the monojet
signature \cite{Bhattacherjee:2013wna,Han:2013usa,Baer:2014cua,Han:2014kaa}. Another interesting possibility is to use the process of vector boson fusion
to search for near-degenerate electroweak partners; this is discussed in \cite{Delannoy:2013ata}. This is a
rapidly evolving area of active investigation and  conclusions from these
studies vary in quantitative detail.  Qualitatively, however, all
studies indicate that probing near-degenerate Higgsinos at the LHC
will remain challenging into the future.

\subsection{Conclusions}

The most naive expectations from naturalness have been dashed by the
results from the Run I searches at the LHC, which so far find no BSM
physics.  In this article, we have summarized the current status of
searches from ATLAS and CMS, with a focus on searches motivated by
solutions to the hierarchy problem and searches for dark matter.  A
summary of these searches is shown in Tables~\ref{tab:compressed}
and~\ref{tab:b2gbh}.

The stringent limits on the existence of new physics at the LHC has
led some to question the utility of naturalness as a guiding principle
to BSM physics.  At one extreme is the camp that is ready to abandon
naturalness altogether; this camp is aided by the observation of what
is consistent with a non-zero, but small, cosmological constant \footnote{See
for example \cite{Weinberg:2012es} for a recent review.} which implies
a fine-tuning of 120 orders of magnitude \cite{Weinberg:1988cp} that
is currently not understood. A less extreme view holds that some level
of fine-tuning seems to be implied by the LHC results, but that
naturalness is still useful to define the next energy scale
milestones; in this view there is still much room to cover before we
are faced with the full fine-tuning of 32 orders of magnitude that
would be implied if no BSM physics appears before the Planck scale.

In our view it is still too early to worry that naturalness is in
trouble.  As described above, there remain broad categories of
topologies without appreciable fine-tuning that have evaded searches
so far.  Furthermore, it is useful to be reminded that naturalness
itself is not well-defined. In addition to the obviously subjective
choice of the level of fine-tuning that one calls unnatural, there are
difficulties quantifying fine-tuning in the first place. A recent
review of the issues can be found in Ref. \cite{Feng:2013pwa}.  At its
core, the naturalness principle was only ever meant to be a rule of
thumb, and as applied to the EW hierarchy, merely a rough expectation
that there should be new physics somewhere around TeV scale. As we
have seen, there are many well-motivated models where particles below
a TeV are still allowed by the LHC.  Indeed, in one extreme example,
the LHC cannot even improve on LEP bounds for a Higgsino LSP, whose
mass plays a key role in natural SUSY.  Given these considerations,
the story of searches for new physics at the energy frontier, still
motivated by the hierarchy problem and by the existence of dark
matter, remains compelling going into the future running of the LHC.

\section*{Acknowledgments}

\noindent The authors are grateful to J.Boyd, M. D'Onofrio, and S. Rappoccio for
their comments on this manuscript.  The authors also thank C. Csaki,
J. Terning, and A. Wulzer for their useful discussions.  This work is
supported by National Science Foundation Award 1306801 (E.H.), by US
Department of Energy contract DE-AC02-98CH10886 (G.R.), and by the
Department of Energy Early Career Award DOE-ARRA-SC0003883 and an
Alfred P. Sloan Foundation Fellowship (D.S.).

\bibliographystyle{utphys}
\bibliography{article}

\providecommand{\href}[2]{#2}\begingroup\raggedright\begin{thebibliography}{100}

\bibitem{Evans:2011zzb}
L.~Evans, ``{The Large Hadron Collider},''
\href{http://dx.doi.org/10.1146/annurev-nucl-102010-130438}{{\em
  Ann.Rev.Nucl.Part.Sci.} {\bfseries 61} (2011) 435--466}.

\bibitem{Aad:2008zzm}
{\bfseries ATLAS} Collaboration, G.~Aad {\em et~al.}, ``{The ATLAS Experiment
  at the CERN Large Hadron Collider},''
\href{http://dx.doi.org/10.1088/1748-0221/3/08/S08003}{{\em JINST} {\bfseries
  3} (2008) S08003}.

\bibitem{Chatrchyan:2008aa}
{\bfseries CMS} Collaboration, S.~Chatrchyan {\em et~al.}, ``{The CMS
  experiment at the CERN LHC},''
\href{http://dx.doi.org/10.1088/1748-0221/3/08/S08004}{{\em JINST} {\bfseries
  3} (2008) S08004}.

\bibitem{Aad:2012tfa}
{\bfseries ATLAS} Collaboration, G.~Aad {\em et~al.}, ``{Observation of a new
  particle in the search for the Standard Model Higgs boson with the ATLAS
  detector at the LHC},''
  \href{http://dx.doi.org/10.1016/j.physletb.2012.08.020}{{\em Phys.Lett.}
  {\bfseries B716} (2012) 1--29},
\href{http://arxiv.org/abs/1207.7214}{{\ttfamily arXiv:1207.7214 [hep-ex]}}.

\bibitem{Chatrchyan:2012ufa}
{\bfseries CMS} Collaboration, S.~Chatrchyan {\em et~al.}, ``{Observation of a
  new boson at a mass of 125 GeV with the CMS experiment at the LHC},''
  \href{http://dx.doi.org/10.1016/j.physletb.2012.08.021}{{\em Phys.Lett.}
  {\bfseries B716} (2012) 30--61},
\href{http://arxiv.org/abs/1207.7235}{{\ttfamily arXiv:1207.7235 [hep-ex]}}.

\bibitem{Weinberg:1975gm}
S.~Weinberg, ``{Implications of Dynamical Symmetry Breaking},''
\href{http://dx.doi.org/10.1103/PhysRevD.13.974}{{\em Phys. Rev.} {\bfseries
  D13} (1976) 974}.

\bibitem{Gildener:1976ai}
E.~Gildener, ``{Gauge Symmetry Hierarchies},''
\href{http://dx.doi.org/10.1103/PhysRevD.14.1667}{{\em Phys. Rev.} {\bfseries
  D14} (1976) 1667}.

\bibitem{Weinberg:1979bn}
S.~Weinberg, ``{Implications of Dynamical Symmetry Breaking: An Addendum},''
\href{http://dx.doi.org/10.1103/PhysRevD.19.1277}{{\em Phys. Rev.} {\bfseries
  D19} (1979) 1277}.

\bibitem{Susskind:1978ms}
L.~Susskind, ``{Dynamics of Spontaneous Symmetry Breaking in the Weinberg-
  Salam Theory},''
\href{http://dx.doi.org/10.1103/PhysRevD.20.2619}{{\em Phys. Rev.} {\bfseries
  D20} (1979) 2619}.

\bibitem{'tHooft:1979bh}
G.~'t~Hooft, ``{Naturalness, chiral symmetry, and spontaneous chiral symmetry
  breaking},''
{\em NATO Adv.Study Inst.Ser.B Phys.} {\bfseries 59} (1980) 135.

\bibitem{Bertone:2004pz}
G.~Bertone, D.~Hooper, and J.~Silk, ``{Particle dark matter: Evidence,
  candidates and constraints},''
  \href{http://dx.doi.org/10.1016/j.physrep.2004.08.031}{{\em Phys.Rept.}
  {\bfseries 405} (2005) 279--390},
\href{http://arxiv.org/abs/hep-ph/0404175}{{\ttfamily arXiv:hep-ph/0404175
  [hep-ph]}}.

\bibitem{Feng:2010gw}
J.~L. Feng, ``{Dark Matter Candidates from Particle Physics and Methods of
  Detection},''
  \href{http://dx.doi.org/10.1146/annurev-astro-082708-101659}{{\em
  Ann.Rev.Astron.Astrophys.} {\bfseries 48} (2010) 495--545},
\href{http://arxiv.org/abs/1003.0904}{{\ttfamily arXiv:1003.0904
  [astro-ph.CO]}}.

\bibitem{Martin:1997ns}
S.~P. Martin, ``{A Supersymmetry primer},''
\href{http://arxiv.org/abs/hep-ph/9709356}{{\ttfamily arXiv:hep-ph/9709356
  [hep-ph]}}.

\bibitem{Beenakker:2011fu}
W.~Beenakker, S.~Brensing, M.~Kramer, A.~Kulesza, E.~Laenen, {\em et~al.},
  ``{Squark and Gluino Hadroproduction},''
  \href{http://dx.doi.org/10.1142/S0217751X11053560}{{\em Int.J.Mod.Phys.}
  {\bfseries A26} (2011) 2637--2664},
\href{http://arxiv.org/abs/1105.1110}{{\ttfamily arXiv:1105.1110 [hep-ph]}}.

\bibitem{Beenakker:1996ed}
W.~Beenakker, R.~Hopker, and M.~Spira, ``{PROSPINO: A Program for the
  production of supersymmetric particles in next-to-leading order QCD},''
\href{http://arxiv.org/abs/hep-ph/9611232}{{\ttfamily arXiv:hep-ph/9611232
  [hep-ph]}}.

\bibitem{Barbier:2004ez}
R.~Barbier, C.~Berat, M.~Besancon, M.~Chemtob, A.~Deandrea, {\em et~al.},
  ``{R-parity violating supersymmetry},''
  \href{http://dx.doi.org/10.1016/j.physrep.2005.08.006}{{\em Phys.Rept.}
  {\bfseries 420} (2005) 1--202},
\href{http://arxiv.org/abs/hep-ph/0406039}{{\ttfamily arXiv:hep-ph/0406039
  [hep-ph]}}.

\bibitem{Giudice:1998bp}
G.~Giudice and R.~Rattazzi, ``{Theories with gauge mediated supersymmetry
  breaking},'' \href{http://dx.doi.org/10.1016/S0370-1573(99)00042-3}{{\em
  Phys.Rept.} {\bfseries 322} (1999) 419--499},
\href{http://arxiv.org/abs/hep-ph/9801271}{{\ttfamily arXiv:hep-ph/9801271
  [hep-ph]}}.

\bibitem{Dine:1993yw}
M.~Dine and A.~E. Nelson, ``{Dynamical supersymmetry breaking at
  low-energies},'' \href{http://dx.doi.org/10.1103/PhysRevD.48.1277}{{\em
  Phys.Rev.} {\bfseries D48} (1993) 1277--1287},
\href{http://arxiv.org/abs/hep-ph/9303230}{{\ttfamily arXiv:hep-ph/9303230
  [hep-ph]}}.

\bibitem{Dine:1994vc}
M.~Dine, A.~E. Nelson, and Y.~Shirman, ``{Low-energy dynamical supersymmetry
  breaking simplified},''
  \href{http://dx.doi.org/10.1103/PhysRevD.51.1362}{{\em Phys.Rev.} {\bfseries
  D51} (1995) 1362--1370},
\href{http://arxiv.org/abs/hep-ph/9408384}{{\ttfamily arXiv:hep-ph/9408384
  [hep-ph]}}.

\bibitem{Dine:1995ag}
M.~Dine, A.~E. Nelson, Y.~Nir, and Y.~Shirman, ``{New tools for low-energy
  dynamical supersymmetry breaking},''
  \href{http://dx.doi.org/10.1103/PhysRevD.53.2658}{{\em Phys.Rev.} {\bfseries
  D53} (1996) 2658--2669},
\href{http://arxiv.org/abs/hep-ph/9507378}{{\ttfamily arXiv:hep-ph/9507378
  [hep-ph]}}.

\bibitem{Meade:2008wd}
P.~Meade, N.~Seiberg, and D.~Shih, ``{General Gauge Mediation},''
  \href{http://dx.doi.org/10.1143/PTPS.177.143}{{\em Prog.Theor.Phys.Suppl.}
  {\bfseries 177} (2009) 143--158},
\href{http://arxiv.org/abs/0801.3278}{{\ttfamily arXiv:0801.3278 [hep-ph]}}.

\bibitem{Buican:2008ws}
M.~Buican, P.~Meade, N.~Seiberg, and D.~Shih, ``{Exploring General Gauge
  Mediation},'' \href{http://dx.doi.org/10.1088/1126-6708/2009/03/016}{{\em
  JHEP} {\bfseries 0903} (2009) 016},
\href{http://arxiv.org/abs/0812.3668}{{\ttfamily arXiv:0812.3668 [hep-ph]}}.

\bibitem{Chamseddine:1982jx}
A.~H. Chamseddine, R.~L. Arnowitt, and P.~Nath, ``{Locally Supersymmetric Grand
  Unification},''
\href{http://dx.doi.org/10.1103/PhysRevLett.49.970}{{\em Phys.Rev.Lett.}
  {\bfseries 49} (1982) 970}.

\bibitem{Barbieri:1982eh}
R.~Barbieri, S.~Ferrara, and C.~A. Savoy, ``{Gauge Models with Spontaneously
  Broken Local Supersymmetry},''
\href{http://dx.doi.org/10.1016/0370-2693(82)90685-2}{{\em Phys.Lett.}
  {\bfseries B119} (1982) 343}.

\bibitem{Ibanez:1982ee}
L.~E. Ibanez, ``{Locally Supersymmetric SU(5) Grand Unification},''
\href{http://dx.doi.org/10.1016/0370-2693(82)90604-9}{{\em Phys.Lett.}
  {\bfseries B118} (1982) 73}.

\bibitem{Hall:1983iz}
L.~J. Hall, J.~D. Lykken, and S.~Weinberg, ``{Supergravity as the Messenger of
  Supersymmetry Breaking},''
\href{http://dx.doi.org/10.1103/PhysRevD.27.2359}{{\em Phys.Rev.} {\bfseries
  D27} (1983) 2359--2378}.

\bibitem{Ohta:1982wn}
N.~Ohta, ``{GRAND UNIFIED THEORIES BASED ON LOCAL SUPERSYMMETRY},''
\href{http://dx.doi.org/10.1143/PTP.70.542}{{\em Prog.Theor.Phys.} {\bfseries
  70} (1983) 542}.

\bibitem{Kane:1993td}
G.~L. Kane, C.~F. Kolda, L.~Roszkowski, and J.~D. Wells, ``{Study of
  constrained minimal supersymmetry},''
  \href{http://dx.doi.org/10.1103/PhysRevD.49.6173}{{\em Phys.Rev.} {\bfseries
  D49} (1994) 6173--6210},
\href{http://arxiv.org/abs/hep-ph/9312272}{{\ttfamily arXiv:hep-ph/9312272
  [hep-ph]}}.

\bibitem{Giudice:1998xp}
G.~F. Giudice, M.~A. Luty, H.~Murayama, and R.~Rattazzi, ``{Gaugino mass
  without singlets},''
  \href{http://dx.doi.org/10.1088/1126-6708/1998/12/027}{{\em JHEP} {\bfseries
  9812} (1998) 027},
\href{http://arxiv.org/abs/hep-ph/9810442}{{\ttfamily arXiv:hep-ph/9810442
  [hep-ph]}}.

\bibitem{Randall:1998uk}
L.~Randall and R.~Sundrum, ``{Out of this world supersymmetry breaking},''
  \href{http://dx.doi.org/10.1016/S0550-3213(99)00359-4}{{\em Nucl.Phys.}
  {\bfseries B557} (1999) 79--118},
\href{http://arxiv.org/abs/hep-th/9810155}{{\ttfamily arXiv:hep-th/9810155
  [hep-th]}}.

\bibitem{Alwall:2008ve}
J.~Alwall, M.-P. Le, M.~Lisanti, and J.~G. Wacker, ``{Searching for Directly
  Decaying Gluinos at the Tevatron},''
  \href{http://dx.doi.org/10.1016/j.physletb.2008.06.065}{{\em Phys.Lett.}
  {\bfseries B666} (2008) 34--37},
\href{http://arxiv.org/abs/0803.0019}{{\ttfamily arXiv:0803.0019 [hep-ph]}}.

\bibitem{Alwall:2008va}
J.~Alwall, M.-P. Le, M.~Lisanti, and J.~G. Wacker, ``{Model-Independent Jets
  plus Missing Energy Searches},''
  \href{http://dx.doi.org/10.1103/PhysRevD.79.015005}{{\em Phys.Rev.}
  {\bfseries D79} (2009) 015005},
\href{http://arxiv.org/abs/0809.3264}{{\ttfamily arXiv:0809.3264 [hep-ph]}}.

\bibitem{Alwall:2008ag}
J.~Alwall, P.~Schuster, and N.~Toro, ``{Simplified Models for a First
  Characterization of New Physics at the LHC},''
  \href{http://dx.doi.org/10.1103/PhysRevD.79.075020}{{\em Phys.Rev.}
  {\bfseries D79} (2009) 075020},
\href{http://arxiv.org/abs/0810.3921}{{\ttfamily arXiv:0810.3921 [hep-ph]}}.

\bibitem{Alves:2011wf}
{\bfseries LHC New Physics Working Group} Collaboration, D.~Alves {\em et~al.},
  ``{Simplified Models for LHC New Physics Searches},''
  \href{http://dx.doi.org/10.1088/0954-3899/39/10/105005}{{\em J.Phys.}
  {\bfseries G39} (2012) 105005},
\href{http://arxiv.org/abs/1105.2838}{{\ttfamily arXiv:1105.2838 [hep-ph]}}.

\bibitem{Feng:2013pwa}
J.~L. Feng, ``{Naturalness and the Status of Supersymmetry},''
  \href{http://dx.doi.org/10.1146/annurev-nucl-102010-130447}{{\em
  Ann.Rev.Nucl.Part.Sci.} {\bfseries 63} (2013) 351--382},
\href{http://arxiv.org/abs/1302.6587}{{\ttfamily arXiv:1302.6587 [hep-ph]}}.

\bibitem{Craig:2013cxa}
N.~Craig, ``{The State of Supersymmetry after Run I of the LHC},''
\href{http://arxiv.org/abs/1309.0528}{{\ttfamily arXiv:1309.0528 [hep-ph]}}.

\bibitem{Bellazzini:2014yua}
B.~Bellazzini, C.~Cs‡ki, and J.~Serra, ``{Composite Higgses},''
  \href{http://dx.doi.org/10.1140/epjc/s10052-014-2766-x}{{\em Eur.Phys.J.}
  {\bfseries C74} (2014) 2766},
\href{http://arxiv.org/abs/1401.2457}{{\ttfamily arXiv:1401.2457 [hep-ph]}}.

\bibitem{Agashe:2004rs}
K.~Agashe, R.~Contino, and A.~Pomarol, ``{The Minimal composite Higgs model},''
  \href{http://dx.doi.org/10.1016/j.nuclphysb.2005.04.035}{{\em Nucl.Phys.}
  {\bfseries B719} (2005) 165--187},
\href{http://arxiv.org/abs/hep-ph/0412089}{{\ttfamily arXiv:hep-ph/0412089
  [hep-ph]}}.

\bibitem{Contino:2010rs}
R.~Contino, ``{The Higgs as a Composite Nambu-Goldstone Boson},''
\href{http://arxiv.org/abs/1005.4269}{{\ttfamily arXiv:1005.4269 [hep-ph]}}.

\bibitem{DeSimone:2012fs}
A.~De~Simone, O.~Matsedonskyi, R.~Rattazzi, and A.~Wulzer, ``{A First Top
  Partner Hunter's Guide},''
  \href{http://dx.doi.org/10.1007/JHEP04(2013)004}{{\em JHEP} {\bfseries 1304}
  (2013) 004},
\href{http://arxiv.org/abs/1211.5663}{{\ttfamily arXiv:1211.5663 [hep-ph]}}.

\bibitem{Aguilar-Saavedra:2013qpa}
J.~Aguilar-Saavedra, R.~Benbrik, S.~Heinemeyer, and M.~Perez-Victoria, ``{A
  handbook of vector-like quarks: mixing and single production},''
  \href{http://dx.doi.org/10.1103/PhysRevD.88.094010}{{\em Phys.Rev.}
  {\bfseries D88} (2013) 094010},
\href{http://arxiv.org/abs/1306.0572}{{\ttfamily arXiv:1306.0572 [hep-ph]}}.

\bibitem{Aliev:2010zk}
M.~Aliev, H.~Lacker, U.~Langenfeld, S.~Moch, P.~Uwer, {\em et~al.}, ``{HATHOR:
  HAdronic Top and Heavy quarks crOss section calculatoR},''
  \href{http://dx.doi.org/10.1016/j.cpc.2010.12.040}{{\em Comput.Phys.Commun.}
  {\bfseries 182} (2011) 1034--1046},
\href{http://arxiv.org/abs/1007.1327}{{\ttfamily arXiv:1007.1327 [hep-ph]}}.

\bibitem{Verlinde:1999fy}
H.~L. Verlinde, ``{Holography and compactification},''
  \href{http://dx.doi.org/10.1016/S0550-3213(00)00224-8}{{\em Nucl.Phys.}
  {\bfseries B580} (2000) 264--274},
\href{http://arxiv.org/abs/hep-th/9906182}{{\ttfamily arXiv:hep-th/9906182
  [hep-th]}}.

\bibitem{Gubser:1999vj}
S.~S. Gubser, ``{AdS / CFT and gravity},''
  \href{http://dx.doi.org/10.1103/PhysRevD.63.084017}{{\em Phys.Rev.}
  {\bfseries D63} (2001) 084017},
\href{http://arxiv.org/abs/hep-th/9912001}{{\ttfamily arXiv:hep-th/9912001
  [hep-th]}}.

\bibitem{Verlinde:2000px}
H.~L. Verlinde, ``{Supersymmetry at large distance scales},''
\href{http://arxiv.org/abs/hep-th/0004003}{{\ttfamily arXiv:hep-th/0004003
  [hep-th]}}.

\bibitem{ArkaniHamed:2000ds}
N.~Arkani-Hamed, M.~Porrati, and L.~Randall, ``{Holography and
  phenomenology},'' \href{http://dx.doi.org/10.1088/1126-6708/2001/08/017}{{\em
  JHEP} {\bfseries 0108} (2001) 017},
\href{http://arxiv.org/abs/hep-th/0012148}{{\ttfamily arXiv:hep-th/0012148
  [hep-th]}}.

\bibitem{Rattazzi:2000hs}
R.~Rattazzi and A.~Zaffaroni, ``{Comments on the holographic picture of the
  Randall-Sundrum model},''
  \href{http://dx.doi.org/10.1088/1126-6708/2001/04/021}{{\em JHEP} {\bfseries
  0104} (2001) 021},
\href{http://arxiv.org/abs/hep-th/0012248}{{\ttfamily arXiv:hep-th/0012248
  [hep-th]}}.

\bibitem{Randall:1999ee}
L.~Randall and R.~Sundrum, ``{A Large mass hierarchy from a small extra
  dimension},'' \href{http://dx.doi.org/10.1103/PhysRevLett.83.3370}{{\em
  Phys.Rev.Lett.} {\bfseries 83} (1999) 3370--3373},
\href{http://arxiv.org/abs/hep-ph/9905221}{{\ttfamily arXiv:hep-ph/9905221
  [hep-ph]}}.

\bibitem{Maldacena:1997re}
J.~M. Maldacena, ``{The Large N limit of superconformal field theories and
  supergravity},'' {\em Adv.Theor.Math.Phys.} {\bfseries 2} (1998) 231--252,
\href{http://arxiv.org/abs/hep-th/9711200}{{\ttfamily arXiv:hep-th/9711200
  [hep-th]}}.

\bibitem{Gubser:1998bc}
S.~Gubser, I.~R. Klebanov, and A.~M. Polyakov, ``{Gauge theory correlators from
  noncritical string theory},''
  \href{http://dx.doi.org/10.1016/S0370-2693(98)00377-3}{{\em Phys.Lett.}
  {\bfseries B428} (1998) 105--114},
\href{http://arxiv.org/abs/hep-th/9802109}{{\ttfamily arXiv:hep-th/9802109
  [hep-th]}}.

\bibitem{Witten:1998qj}
E.~Witten, ``{Anti-de Sitter space and holography},'' {\em
  Adv.Theor.Math.Phys.} {\bfseries 2} (1998) 253--291,
\href{http://arxiv.org/abs/hep-th/9802150}{{\ttfamily arXiv:hep-th/9802150
  [hep-th]}}.

\bibitem{Contino:2003ve}
R.~Contino, Y.~Nomura, and A.~Pomarol, ``{Higgs as a holographic
  pseudoGoldstone boson},''
  \href{http://dx.doi.org/10.1016/j.nuclphysb.2003.08.027}{{\em Nucl.Phys.}
  {\bfseries B671} (2003) 148--174},
\href{http://arxiv.org/abs/hep-ph/0306259}{{\ttfamily arXiv:hep-ph/0306259
  [hep-ph]}}.

\bibitem{Beltran:2010ww}
M.~Beltran, D.~Hooper, E.~W. Kolb, Z.~A. Krusberg, and T.~M. Tait, ``{Maverick
  dark matter at colliders},''
  \href{http://dx.doi.org/10.1007/JHEP09(2010)037}{{\em JHEP} {\bfseries 1009}
  (2010) 037},
\href{http://arxiv.org/abs/1002.4137}{{\ttfamily arXiv:1002.4137 [hep-ph]}}.

\bibitem{Rajaraman:2011wf}
A.~Rajaraman, W.~Shepherd, T.~M. Tait, and A.~M. Wijangco, ``{LHC Bounds on
  Interactions of Dark Matter},''
  \href{http://dx.doi.org/10.1103/PhysRevD.84.095013}{{\em Phys.Rev.}
  {\bfseries D84} (2011) 095013},
\href{http://arxiv.org/abs/1108.1196}{{\ttfamily arXiv:1108.1196 [hep-ph]}}.

\bibitem{Fox:2011pm}
P.~J. Fox, R.~Harnik, J.~Kopp, and Y.~Tsai, ``{Missing Energy Signatures of
  Dark Matter at the LHC},''
  \href{http://dx.doi.org/10.1103/PhysRevD.85.056011}{{\em Phys.Rev.}
  {\bfseries D85} (2012) 056011},
\href{http://arxiv.org/abs/1109.4398}{{\ttfamily arXiv:1109.4398 [hep-ph]}}.

\bibitem{Shoemaker:2011vi}
I.~M. Shoemaker and L.~Vecchi, ``{Unitarity and Monojet Bounds on Models for
  DAMA, CoGeNT, and CRESST-II},''
  \href{http://dx.doi.org/10.1103/PhysRevD.86.015023}{{\em Phys.Rev.}
  {\bfseries D86} (2012) 015023},
\href{http://arxiv.org/abs/1112.5457}{{\ttfamily arXiv:1112.5457 [hep-ph]}}.

\bibitem{Busoni:2013lha}
G.~Busoni, A.~De~Simone, E.~Morgante, and A.~Riotto, ``{On the Validity of the
  Effective Field Theory for Dark Matter Searches at the LHC},''
  \href{http://dx.doi.org/10.1016/j.physletb.2013.11.069}{{\em Phys.Lett.}
  {\bfseries B728} (2014) 412--421},
\href{http://arxiv.org/abs/1307.2253}{{\ttfamily arXiv:1307.2253 [hep-ph]}}.

\bibitem{Buchmueller:2013dya}
O.~Buchmueller, M.~J. Dolan, and C.~McCabe, ``{Beyond Effective Field Theory
  for Dark Matter Searches at the LHC},''
  \href{http://dx.doi.org/10.1007/JHEP01(2014)025}{{\em JHEP} {\bfseries 1401}
  (2014) 025},
\href{http://arxiv.org/abs/1308.6799}{{\ttfamily arXiv:1308.6799 [hep-ph]}}.

\bibitem{ATLASPublic}
{http://twiki.cern.ch/twiki/bin/view/AtlasPublic}.

\bibitem{CMSPublic}
{http://twiki.cern.ch/twiki/bin/view/CMSPublic/PhysicsResults}.

\bibitem{Chatrchyan:2014lfa}
{\bfseries CMS Collaboration} Collaboration, S.~Chatrchyan {\em et~al.},
  ``{Search for new physics in the multijet and missing transverse momentum
  final state in proton-proton collisions at $\sqrt{s}$= 8 TeV},''
  \href{http://dx.doi.org/10.1007/JHEP06(2014)055}{{\em JHEP} {\bfseries 1406}
  (2014) 055},
\href{http://arxiv.org/abs/1402.4770}{{\ttfamily arXiv:1402.4770 [hep-ex]}}.

\bibitem{Aad:2013wta}
{\bfseries ATLAS} Collaboration, G.~Aad {\em et~al.}, ``{Search for new
  phenomena in final states with large jet multiplicities and missing
  transverse momentum at sqrt(s)=8 TeV proton-proton collisions using the ATLAS
  experiment},'' \href{http://dx.doi.org/10.1007/JHEP10(2013)130}{{\em JHEP}
  {\bfseries 1310} (2013) 130},
\href{http://arxiv.org/abs/1308.1841}{{\ttfamily arXiv:1308.1841 [hep-ex]}}.

\bibitem{Chatrchyan:2013fea}
{\bfseries CMS} Collaboration, S.~Chatrchyan {\em et~al.}, ``{Search for new
  physics in events with same-sign dileptons and jets in pp collisions at
  $\sqrt{s}$=8 TeV},'' \href{http://dx.doi.org/10.1007/JHEP01(2014)163}{{\em
  JHEP} {\bfseries 1401} (2014) 163},
\href{http://arxiv.org/abs/1311.6736}{{\ttfamily arXiv:1311.6736 [hep-ex]}}.

\bibitem{Chatrchyan:2013lya}
{\bfseries CMS} Collaboration, S.~Chatrchyan {\em et~al.}, ``{Search for
  supersymmetry in hadronic final states with missing transverse energy using
  the variables AlphaT and b-quark multiplicity in pp collisions at 8 TeV},''
  \href{http://dx.doi.org/10.1140/epjc/s10052-013-2568-6}{{\em Eur.Phys.J.}
  {\bfseries C73} (2013) 2568},
\href{http://arxiv.org/abs/1303.2985}{{\ttfamily arXiv:1303.2985 [hep-ex]}}.

\bibitem{Chatrchyan:2013iqa}
{\bfseries CMS Collaboration} Collaboration, S.~Chatrchyan {\em et~al.},
  ``{Search for supersymmetry in pp collisions at $\sqrt{s}$=8 TeV in events
  with a single lepton, large jet multiplicity, and multiple b jets},''
  \href{http://dx.doi.org/10.1016/j.physletb.2014.04.023}{{\em Phys.Lett.}
  {\bfseries B733} (2014) 328--353},
\href{http://arxiv.org/abs/1311.4937}{{\ttfamily arXiv:1311.4937 [hep-ex]}}.

\bibitem{Chatrchyan:2013wxa}
{\bfseries CMS} Collaboration, S.~Chatrchyan {\em et~al.}, ``{Search for gluino
  mediated bottom- and top-squark production in multijet final states in pp
  collisions at 8 TeV},''
  \href{http://dx.doi.org/10.1016/j.physletb.2013.06.058}{{\em Phys.Lett.}
  {\bfseries B725} (2013) 243--270},
\href{http://arxiv.org/abs/1305.2390}{{\ttfamily arXiv:1305.2390 [hep-ex]}}.

\bibitem{Chatrchyan:2014aea}
{\bfseries CMS Collaboration} Collaboration, S.~Chatrchyan {\em et~al.},
  ``{Search for anomalous production of events with three or more leptons in pp
  collisions at $\sqrt{s}$=8 TeV},''
\href{http://arxiv.org/abs/1404.5801}{{\ttfamily arXiv:1404.5801 [hep-ex]}}.

\bibitem{Chatrchyan:2013xna}
{\bfseries CMS} Collaboration, S.~Chatrchyan {\em et~al.}, ``{Search for
  top-squark pair production in the single-lepton final state in pp collisions
  at $\sqrt{s}$ = 8 TeV},''
  \href{http://dx.doi.org/10.1140/epjc/s10052-013-2677-2}{{\em Eur.Phys.J.}
  {\bfseries C73} (2013) 2677},
\href{http://arxiv.org/abs/1308.1586}{{\ttfamily arXiv:1308.1586 [hep-ex]}}.

\bibitem{Aad:2013ija}
{\bfseries ATLAS} Collaboration, G.~Aad {\em et~al.}, ``{Search for direct
  third-generation squark pair production in final states with missing
  transverse momentum and two $b$-jets in $\sqrt{s} =$ 8 TeV $pp$ collisions
  with the ATLAS detector},''
  \href{http://dx.doi.org/10.1007/JHEP10(2013)189}{{\em JHEP} {\bfseries 1310}
  (2013) 189},
\href{http://arxiv.org/abs/1308.2631}{{\ttfamily arXiv:1308.2631 [hep-ex]}}.

\bibitem{Chatrchyan:2013mya}
{\bfseries CMS Collaboration} Collaboration, S.~Chatrchyan {\em et~al.},
  ``{Search for top squark and higgsino production using diphoton Higgs boson
  decays},'' \href{http://dx.doi.org/10.1103/PhysRevLett.112.161802}{{\em
  Phys.Rev.Lett.} {\bfseries 112} (2014) 161802},
\href{http://arxiv.org/abs/1312.3310}{{\ttfamily arXiv:1312.3310 [hep-ex]}}.

\bibitem{Aad:2014nua}
{\bfseries ATLAS Collaboration} Collaboration, G.~Aad {\em et~al.}, ``{Search
  for direct production of charginos and neutralinos in events with three
  leptons and missing transverse momentum in $\sqrt{s} =$ 8TeV $pp$ collisions
  with the ATLAS detector},''
  \href{http://dx.doi.org/10.1007/JHEP04(2014)169}{{\em JHEP} {\bfseries 1404}
  (2014) 169},
\href{http://arxiv.org/abs/1402.7029}{{\ttfamily arXiv:1402.7029 [hep-ex]}}.

\bibitem{Djouadi:1998di}
{\bfseries MSSM Working Group} Collaboration, A.~Djouadi {\em et~al.}, ``{The
  Minimal supersymmetric standard model: Group summary report},''
\href{http://arxiv.org/abs/hep-ph/9901246}{{\ttfamily arXiv:hep-ph/9901246
  [hep-ph]}}.

\bibitem{Farrar:1978xj}
G.~R. Farrar and P.~Fayet, ``{Phenomenology of the Production, Decay, and
  Detection of New Hadronic States Associated with Supersymmetry},''
\href{http://dx.doi.org/10.1016/0370-2693(78)90858-4}{{\em Phys. Lett.}
  {\bfseries B76} (1978) 575}.

\bibitem{ArkaniHamed:2004fb}
N.~Arkani-Hamed and S.~Dimopoulos, ``{Supersymmetric unification without low
  energy supersymmetry and signatures for fine-tuning at the LHC},''
  \href{http://dx.doi.org/10.1088/1126-6708/2005/06/073}{{\em JHEP} {\bfseries
  0506} (2005) 073},
\href{http://arxiv.org/abs/hep-th/0405159}{{\ttfamily arXiv:hep-th/0405159
  [hep-th]}}.

\bibitem{ArkaniHamed:2004yi}
N.~Arkani-Hamed, S.~Dimopoulos, G.~Giudice, and A.~Romanino, ``{Aspects of
  split supersymmetry},''
  \href{http://dx.doi.org/10.1016/j.nuclphysb.2004.12.026}{{\em Nucl.Phys.}
  {\bfseries B709} (2005) 3--46},
\href{http://arxiv.org/abs/hep-ph/0409232}{{\ttfamily arXiv:hep-ph/0409232
  [hep-ph]}}.

\bibitem{Fairbairn:2006gg}
M.~Fairbairn, A.~Kraan, D.~Milstead, T.~Sjostrand, P.~Z. Skands, {\em et~al.},
  ``{Stable massive particles at colliders},''
  \href{http://dx.doi.org/10.1016/j.physrep.2006.10.002}{{\em Phys.Rept.}
  {\bfseries 438} (2007) 1--63},
\href{http://arxiv.org/abs/hep-ph/0611040}{{\ttfamily arXiv:hep-ph/0611040
  [hep-ph]}}.

\bibitem{Raklev:2009mg}
A.~R. Raklev, ``{Massive Metastable Charged (S)Particles at the LHC},''
  \href{http://dx.doi.org/10.1142/S0217732309031648}{{\em Mod.Phys.Lett.}
  {\bfseries A24} (2009) 1955--1969},
\href{http://arxiv.org/abs/0908.0315}{{\ttfamily arXiv:0908.0315 [hep-ph]}}.

\bibitem{Kraan:2004tz}
A.~C. Kraan, ``{Interactions of heavy stable hadronizing particles},''
  \href{http://dx.doi.org/10.1140/epjc/s2004-01997-7}{{\em Eur.Phys.J.}
  {\bfseries C37} (2004) 91--104},
\href{http://arxiv.org/abs/hep-ex/0404001}{{\ttfamily arXiv:hep-ex/0404001
  [hep-ex]}}.

\bibitem{Mackeprang:2006gx}
R.~Mackeprang and A.~Rizzi, ``{Interactions of Coloured Heavy Stable Particles
  in Matter},'' \href{http://dx.doi.org/10.1140/epjc/s10052-007-0252-4}{{\em
  Eur.Phys.J.} {\bfseries C50} (2007) 353--362},
\href{http://arxiv.org/abs/hep-ph/0612161}{{\ttfamily arXiv:hep-ph/0612161
  [hep-ph]}}.

\bibitem{Mackeprang:2009ad}
R.~Mackeprang and D.~Milstead, ``{An Updated Description of Heavy-Hadron
  Interactions in GEANT-4},''
  \href{http://dx.doi.org/10.1140/epjc/s10052-010-1262-1}{{\em Eur.Phys.J.}
  {\bfseries C66} (2010) 493--501},
\href{http://arxiv.org/abs/0908.1868}{{\ttfamily arXiv:0908.1868 [hep-ph]}}.

\bibitem{Chatrchyan:2013oca}
{\bfseries CMS} Collaboration, S.~Chatrchyan {\em et~al.}, ``{Searches for
  long-lived charged particles in pp collisions at $\sqrt{s}$=7 and 8 TeV},''
  \href{http://dx.doi.org/10.1007/JHEP07(2013)122}{{\em JHEP} {\bfseries 1307}
  (2013) 122},
\href{http://arxiv.org/abs/1305.0491}{{\ttfamily arXiv:1305.0491 [hep-ex]}}.

\bibitem{Aad:2013gva}
{\bfseries ATLAS} Collaboration, G.~Aad {\em et~al.}, ``{Search for long-lived
  stopped R-hadrons decaying out-of-time with pp collisions using the ATLAS
  detector},'' \href{http://dx.doi.org/10.1103/PhysRevD.88.112003}{{\em
  Phys.Rev.} {\bfseries D88} (2013) 112003},
\href{http://arxiv.org/abs/1310.6584}{{\ttfamily arXiv:1310.6584 [hep-ex]}}.

\bibitem{Dreiner:2012gx}
H.~K. Dreiner, M.~Kramer, and J.~Tattersall, ``{How low can SUSY go? Matching,
  monojets and compressed spectra},''
  \href{http://dx.doi.org/10.1209/0295-5075/99/61001}{{\em Europhys.Lett.}
  {\bfseries 99} (2012) 61001},
\href{http://arxiv.org/abs/1207.1613}{{\ttfamily arXiv:1207.1613 [hep-ph]}}.

\bibitem{Dreiner:2012sh}
H.~Dreiner, M.~KrŠmer, and J.~Tattersall, ``{Exploring QCD uncertainties when
  setting limits on compressed supersymmetric spectra},''
  \href{http://dx.doi.org/10.1103/PhysRevD.87.035006}{{\em Phys.Rev.}
  {\bfseries D87} no.~3, (2013) 035006},
\href{http://arxiv.org/abs/1211.4981}{{\ttfamily arXiv:1211.4981 [hep-ph]}}.

\bibitem{Aad:2013yna}
{\bfseries ATLAS} Collaboration, G.~Aad {\em et~al.}, ``{Search for charginos
  nearly mass-degenerate with the lightest neutralino based on a
  disappearing-track signature in pp collisions at $\sqrt{s}$ = 8 TeV with the
  ATLAS detector},'' \href{http://dx.doi.org/10.1103/PhysRevD.88.112006}{{\em
  Phys.Rev.} {\bfseries D88} (2013) 112006},
\href{http://arxiv.org/abs/1310.3675}{{\ttfamily arXiv:1310.3675 [hep-ex]}}.

\bibitem{Chatrchyan:2013xsw}
{\bfseries CMS} Collaboration, S.~Chatrchyan {\em et~al.}, ``{Search for top
  squarks in R-parity-violating supersymmetry using three or more leptons and
  b-tagged jets},''
  \href{http://dx.doi.org/10.1103/PhysRevLett.111.221801}{{\em Phys.Rev.Lett.}
  {\bfseries 111} (2013) 221801},
\href{http://arxiv.org/abs/1306.6643}{{\ttfamily arXiv:1306.6643 [hep-ex]}}.

\bibitem{Chatrchyan:2013gia}
{\bfseries CMS} Collaboration, S.~Chatrchyan {\em et~al.}, ``{Searches for
  light- and heavy-flavour three-jet resonances in pp collisions at $\sqrt{s}$
  = 8 TeV},'' \href{http://dx.doi.org/10.1016/j.physletb.2014.01.049}{{\em
  Phys.Lett.} {\bfseries B730} (2014) 193},
\href{http://arxiv.org/abs/1311.1799}{{\ttfamily arXiv:1311.1799 [hep-ex]}}.

\bibitem{Chatrchyan:2013wfa}
{\bfseries CMS Collaboration} Collaboration, S.~Chatrchyan {\em et~al.},
  ``{Search for top-quark partners with charge 5/3 in the same-sign dilepton
  final state},'' \href{http://dx.doi.org/10.1103/PhysRevLett.112.171801}{{\em
  Phys.Rev.Lett.} {\bfseries 112} (2014) 171801},
\href{http://arxiv.org/abs/1312.2391}{{\ttfamily arXiv:1312.2391 [hep-ex]}}.

\bibitem{Chatrchyan:2013uxa}
{\bfseries CMS} Collaboration, S.~Chatrchyan {\em et~al.}, ``{Inclusive search
  for a vector-like T quark with charge 2/3 in pp collisions at sqrt(s) = 8
  TeV},'' \href{http://dx.doi.org/10.1016/j.physletb.2014.01.006}{{\em
  Phys.Lett.} {\bfseries B729} (2014) 149--171},
\href{http://arxiv.org/abs/1311.7667}{{\ttfamily arXiv:1311.7667 [hep-ex]}}.

\bibitem{Chatrchyan:2014koa}
{\bfseries CMS Collaboration} Collaboration, S.~Chatrchyan {\em et~al.},
  ``{Search for W' $\to $ tb decays in the lepton + jets final state in pp
  collisions at $\sqrt{s}$ = 8 TeV},''
  \href{http://dx.doi.org/10.1007/JHEP05(2014)108}{{\em JHEP} {\bfseries 1405}
  (2014) 108},
\href{http://arxiv.org/abs/1402.2176}{{\ttfamily arXiv:1402.2176 [hep-ex]}}.

\bibitem{Chatrchyan:2013oba}
{\bfseries CMS Collaboration} Collaboration, S.~Chatrchyan {\em et~al.},
  ``{Search for pair production of excited top quarks in the lepton + jets
  final state},'' \href{http://dx.doi.org/10.1007/JHEP06(2014)125}{{\em JHEP}
  {\bfseries 1406} (2014) 125},
\href{http://arxiv.org/abs/1311.5357}{{\ttfamily arXiv:1311.5357 [hep-ex]}}.

\bibitem{Chatrchyan:2013lca}
{\bfseries CMS} Collaboration, S.~Chatrchyan {\em et~al.}, ``{Searches for new
  physics using the $t\bar{t}$ invariant mass distribution in pp collisions at
  $\sqrt{s}$=8  TeV},''
  \href{http://dx.doi.org/10.1103/PhysRevLett.111.211804}{{\em Phys.Rev.Lett.}
  {\bfseries 111} (2013) 211804},
\href{http://arxiv.org/abs/1309.2030}{{\ttfamily arXiv:1309.2030 [hep-ex]}}.

\bibitem{Banks:1999gd}
T.~Banks and W.~Fischler, ``{A Model for high-energy scattering in quantum
  gravity},''
\href{http://arxiv.org/abs/hep-th/9906038}{{\ttfamily arXiv:hep-th/9906038
  [hep-th]}}.

\bibitem{Giddings:2001bu}
S.~B. Giddings and S.~D. Thomas, ``{High-energy colliders as black hole
  factories: The End of short distance physics},''
  \href{http://dx.doi.org/10.1103/PhysRevD.65.056010}{{\em Phys.Rev.}
  {\bfseries D65} (2002) 056010},
\href{http://arxiv.org/abs/hep-ph/0106219}{{\ttfamily arXiv:hep-ph/0106219
  [hep-ph]}}.

\bibitem{Dimopoulos:2001hw}
S.~Dimopoulos and G.~L. Landsberg, ``{Black holes at the LHC},''
  \href{http://dx.doi.org/10.1103/PhysRevLett.87.161602}{{\em Phys.Rev.Lett.}
  {\bfseries 87} (2001) 161602},
\href{http://arxiv.org/abs/hep-ph/0106295}{{\ttfamily arXiv:hep-ph/0106295
  [hep-ph]}}.

\bibitem{Chatrchyan:2013xva}
{\bfseries CMS} Collaboration, S.~Chatrchyan {\em et~al.}, ``{Search for
  microscopic black holes in pp collisions at $\sqrt{s}$ = 8 TeV},''
  \href{http://dx.doi.org/10.1007/JHEP07(2013)178}{{\em JHEP} {\bfseries 1307}
  (2013) 178},
\href{http://arxiv.org/abs/1303.5338}{{\ttfamily arXiv:1303.5338 [hep-ex]}}.

\bibitem{Aad:2013gma}
{\bfseries ATLAS Collaboration} Collaboration, G.~Aad {\em et~al.}, ``{Search
  for Quantum Black-Hole Production in High-Invariant-Mass Lepton+Jet Final
  States Using Proton-Proton Collisions at $\sqrt{s} = 8$ TeV and the ATLAS
  Detector},'' \href{http://dx.doi.org/10.1103/PhysRevLett.112.091804}{{\em
  Phys.Rev.Lett.} {\bfseries 112} (2014) 091804},
\href{http://arxiv.org/abs/1311.2006}{{\ttfamily arXiv:1311.2006 [hep-ex]}}.

\bibitem{Aad:2013cva}
{\bfseries ATLAS} Collaboration, G.~Aad {\em et~al.}, ``{Search for new
  phenomena in photon+jet events collected in proton--proton collisions at
  $\sqrt{s}$ = 8 TeV with the ATLAS detector},''
  \href{http://dx.doi.org/10.1016/j.physletb.2013.12.029}{{\em Phys.Lett.}
  {\bfseries B728} (2014) 562--578},
\href{http://arxiv.org/abs/1309.3230}{{\ttfamily arXiv:1309.3230 [hep-ex]}}.

\bibitem{Aad:2013lna}
{\bfseries ATLAS} Collaboration, G.~Aad {\em et~al.}, ``{Search for microscopic
  black holes in a like-sign dimuon final state using large track multiplicity
  with the ATLAS detector},''
  \href{http://dx.doi.org/10.1103/PhysRevD.88.072001}{{\em Phys.Rev.}
  {\bfseries D88} (2013) 072001},
\href{http://arxiv.org/abs/1308.4075}{{\ttfamily arXiv:1308.4075 [hep-ex]}}.

\bibitem{Aad:2013oja}
{\bfseries ATLAS} Collaboration, G.~Aad {\em et~al.}, ``{Search for dark matter
  in events with a hadronically decaying W or Z boson and missing transverse
  momentum in pp collisions at $\sqrt{s}$=8 TeV with the ATLAS detector},''
  \href{http://dx.doi.org/10.1103/PhysRevLett.112.041802}{{\em Phys.Rev.Lett.}
  {\bfseries 112} (2014) 041802},
\href{http://arxiv.org/abs/1309.4017}{{\ttfamily arXiv:1309.4017 [hep-ex]}}.

\bibitem{Trotta:2008bp}
R.~Trotta, F.~Feroz, M.~P. Hobson, L.~Roszkowski, and R.~Ruiz~de Austri, ``{The
  Impact of priors and observables on parameter inferences in the Constrained
  MSSM},'' \href{http://dx.doi.org/10.1088/1126-6708/2008/12/024}{{\em JHEP}
  {\bfseries 0812} (2008) 024},
\href{http://arxiv.org/abs/0809.3792}{{\ttfamily arXiv:0809.3792 [hep-ph]}}.

\bibitem{Buchmueller:2009fn}
O.~Buchmueller, R.~Cavanaugh, A.~De~Roeck, J.~Ellis, H.~Flacher, {\em et~al.},
  ``{Likelihood Functions for Supersymmetric Observables in Frequentist
  Analyses of the CMSSM and NUHM1},''
  \href{http://dx.doi.org/10.1140/epjc/s10052-009-1159-z}{{\em Eur.Phys.J.}
  {\bfseries C64} (2009) 391--415},
\href{http://arxiv.org/abs/0907.5568}{{\ttfamily arXiv:0907.5568 [hep-ph]}}.

\bibitem{Bechtle:2013mda}
P.~Bechtle, K.~Desch, H.~K. Dreiner, M.~Hamer, M.~KrŠmer, {\em et~al.},
  ``{Constrained Supersymmetry after the Higgs Boson Discovery: A global
  analysis with Fittino},''
\href{http://arxiv.org/abs/1310.3045}{{\ttfamily arXiv:1310.3045 [hep-ph]}}.

\bibitem{Buchmueller:2013rsa}
O.~Buchmueller, R.~Cavanaugh, A.~De~Roeck, M.~Dolan, J.~Ellis, {\em et~al.},
  ``{The CMSSM and NUHM1 after LHC Run 1},''
\href{http://arxiv.org/abs/1312.5250}{{\ttfamily arXiv:1312.5250 [hep-ph]}}.

\bibitem{Ajaib:2012vc}
M.~A. Ajaib, I.~Gogoladze, F.~Nasir, and Q.~Shafi, ``{Revisiting mGMSB in Light
  of a 125 GeV Higgs},''
  \href{http://dx.doi.org/10.1016/j.physletb.2012.06.036}{{\em Phys.Lett.}
  {\bfseries B713} (2012) 462--468},
\href{http://arxiv.org/abs/1204.2856}{{\ttfamily arXiv:1204.2856 [hep-ph]}}.

\bibitem{Hall:2011aa}
L.~J. Hall, D.~Pinner, and J.~T. Ruderman, ``{A Natural SUSY Higgs Near 126
  GeV},'' {\em JHEP} {\bfseries 1204} (2012) 131,
\href{http://arxiv.org/abs/1112.2703}{{\ttfamily arXiv:1112.2703 [hep-ph]}}.

\bibitem{Heinemeyer:2011aa}
S.~Heinemeyer, O.~Stal, and G.~Weiglein, ``{Interpreting the LHC Higgs Search
  Results in the MSSM},''
  \href{http://dx.doi.org/10.1016/j.physletb.2012.02.084}{{\em Phys.Lett.}
  {\bfseries B710} (2012) 201--206},
\href{http://arxiv.org/abs/1112.3026}{{\ttfamily arXiv:1112.3026 [hep-ph]}}.

\bibitem{Arbey:2011ab}
A.~Arbey, M.~Battaglia, A.~Djouadi, F.~Mahmoudi, and J.~Quevillon,
  ``{Implications of a 125 GeV Higgs for supersymmetric models},''
  \href{http://dx.doi.org/10.1016/j.physletb.2012.01.053}{{\em Phys.Lett.}
  {\bfseries B708} (2012) 162--169},
\href{http://arxiv.org/abs/1112.3028}{{\ttfamily arXiv:1112.3028 [hep-ph]}}.

\bibitem{Draper:2011aa}
P.~Draper, P.~Meade, M.~Reece, and D.~Shih, ``{Implications of a 125 GeV Higgs
  for the MSSM and Low-Scale SUSY Breaking},''
  \href{http://dx.doi.org/10.1103/PhysRevD.85.095007}{{\em Phys.Rev.}
  {\bfseries D85} (2012) 095007},
\href{http://arxiv.org/abs/1112.3068}{{\ttfamily arXiv:1112.3068 [hep-ph]}}.

\bibitem{Carena:2011aa}
M.~Carena, S.~Gori, N.~R. Shah, and C.~E. Wagner, ``{A 125 GeV SM-like Higgs in
  the MSSM and the $\gamma \gamma$ rate},''
  \href{http://dx.doi.org/10.1007/JHEP03(2012)014}{{\em JHEP} {\bfseries 1203}
  (2012) 014},
\href{http://arxiv.org/abs/1112.3336}{{\ttfamily arXiv:1112.3336 [hep-ph]}}.

\bibitem{Casas:1994us}
J.~Casas, J.~Espinosa, M.~Quiros, and A.~Riotto, ``{The Lightest Higgs boson
  mass in the minimal supersymmetric standard model},''
  \href{http://dx.doi.org/10.1016/0550-3213(94)00508-C}{{\em Nucl.Phys.}
  {\bfseries B436} (1995) 3--29},
\href{http://arxiv.org/abs/hep-ph/9407389}{{\ttfamily arXiv:hep-ph/9407389
  [hep-ph]}}.

\bibitem{Carena:1995bx}
M.~S. Carena, J.~Espinosa, M.~Quiros, and C.~Wagner, ``{Analytical expressions
  for radiatively corrected Higgs masses and couplings in the MSSM},''
  \href{http://dx.doi.org/10.1016/0370-2693(95)00694-G}{{\em Phys.Lett.}
  {\bfseries B355} (1995) 209--221},
\href{http://arxiv.org/abs/hep-ph/9504316}{{\ttfamily arXiv:hep-ph/9504316
  [hep-ph]}}.

\bibitem{Haber:1996fp}
H.~E. Haber, R.~Hempfling, and A.~H. Hoang, ``{Approximating the radiatively
  corrected Higgs mass in the minimal supersymmetric model},''
  \href{http://dx.doi.org/10.1007/s002880050498}{{\em Z.Phys.} {\bfseries C75}
  (1997) 539--554},
\href{http://arxiv.org/abs/hep-ph/9609331}{{\ttfamily arXiv:hep-ph/9609331
  [hep-ph]}}.

\bibitem{Evans:2012hg}
J.~L. Evans, M.~Ibe, S.~Shirai, and T.~T. Yanagida, ``{A 125GeV Higgs Boson and
  Muon g-2 in More Generic Gauge Mediation},''
  \href{http://dx.doi.org/10.1103/PhysRevD.85.095004}{{\em Phys.Rev.}
  {\bfseries D85} (2012) 095004},
\href{http://arxiv.org/abs/1201.2611}{{\ttfamily arXiv:1201.2611 [hep-ph]}}.

\bibitem{Kang:2012ra}
Z.~Kang, T.~Li, T.~Liu, C.~Tong, and J.~M. Yang, ``{A Heavy SM-like Higgs and a
  Light Stop from Yukawa-Deflected Gauge Mediation},''
  \href{http://dx.doi.org/10.1103/PhysRevD.86.095020}{{\em Phys.Rev.}
  {\bfseries D86} (2012) 095020},
\href{http://arxiv.org/abs/1203.2336}{{\ttfamily arXiv:1203.2336 [hep-ph]}}.

\bibitem{Craig:2012xp}
N.~Craig, S.~Knapen, D.~Shih, and Y.~Zhao, ``{A Complete Model of Low-Scale
  Gauge Mediation},'' \href{http://dx.doi.org/10.1007/JHEP03(2013)154}{{\em
  JHEP} {\bfseries 1303} (2013) 154},
\href{http://arxiv.org/abs/1206.4086}{{\ttfamily arXiv:1206.4086 [hep-ph]}}.

\bibitem{Abdullah:2012tq}
M.~Abdullah, I.~Galon, Y.~Shadmi, and Y.~Shirman, ``{Flavored Gauge Mediation,
  A Heavy Higgs, and Supersymmetric Alignment},''
  \href{http://dx.doi.org/10.1007/JHEP06(2013)057}{{\em JHEP} {\bfseries 1306}
  (2013) 057},
\href{http://arxiv.org/abs/1209.4904}{{\ttfamily arXiv:1209.4904 [hep-ph]}}.

\bibitem{Craig:2013wga}
N.~Craig, S.~Knapen, and D.~Shih, ``{General Messenger Higgs Mediation},''
  \href{http://dx.doi.org/10.1007/JHEP08(2013)118}{{\em JHEP} {\bfseries 1308}
  (2013) 118},
\href{http://arxiv.org/abs/1302.2642}{{\ttfamily arXiv:1302.2642}}.

\bibitem{Byakti:2013ti}
P.~Byakti and T.~S. Ray, ``{Burgeoning the Higgs mass to 125 GeV through
  messenger-matter interactions in GMSB models},''
  \href{http://dx.doi.org/10.1007/JHEP05(2013)055}{{\em JHEP} {\bfseries 1305}
  (2013) 055},
\href{http://arxiv.org/abs/1301.7605}{{\ttfamily arXiv:1301.7605 [hep-ph]}}.

\bibitem{Evans:2013kxa}
J.~A. Evans and D.~Shih, ``{Surveying Extended GMSB Models with $m$$_{h}$=125
  GeV},'' \href{http://dx.doi.org/10.1007/JHEP08(2013)093}{{\em JHEP}
  {\bfseries 1308} (2013) 093},
\href{http://arxiv.org/abs/1303.0228}{{\ttfamily arXiv:1303.0228 [hep-ph]}}.

\bibitem{Calibbi:2013mka}
L.~Calibbi, P.~Paradisi, and R.~Ziegler, ``{Gauge Mediation beyond Minimal
  Flavor Violation},'' \href{http://dx.doi.org/10.1007/JHEP06(2013)052}{{\em
  JHEP} {\bfseries 1306} (2013) 052},
\href{http://arxiv.org/abs/1304.1453}{{\ttfamily arXiv:1304.1453 [hep-ph]}}.

\bibitem{Jelinski:2013kta}
T.~Jeli?ski, ``{On messengers couplings in extended GMSB models},''
  \href{http://dx.doi.org/10.1007/JHEP09(2013)107}{{\em JHEP} {\bfseries 1309}
  (2013) 107},
\href{http://arxiv.org/abs/1305.6277}{{\ttfamily arXiv:1305.6277 [hep-ph]}}.

\bibitem{Knapen:2013zla}
S.~Knapen and D.~Shih, ``{Higgs Mediation with Strong Hidden Sector
  Dynamics},''
\href{http://arxiv.org/abs/1311.7107}{{\ttfamily arXiv:1311.7107 [hep-ph]}}.

\bibitem{Kumar:2014uxa}
P.~Kumar, D.~Li, D.~Poland, and A.~Stergiou, ``{OPE Methods for the Holomorphic
  Higgs Portal},''
\href{http://arxiv.org/abs/1401.7690}{{\ttfamily arXiv:1401.7690 [hep-ph]}}.

\bibitem{Ellwanger:2009dp}
U.~Ellwanger, C.~Hugonie, and A.~M. Teixeira, ``{The Next-to-Minimal
  Supersymmetric Standard Model},''
  \href{http://dx.doi.org/10.1016/j.physrep.2010.07.001}{{\em Phys.Rept.}
  {\bfseries 496} (2010) 1--77},
\href{http://arxiv.org/abs/0910.1785}{{\ttfamily arXiv:0910.1785 [hep-ph]}}.

\bibitem{Batra:2003nj}
P.~Batra, A.~Delgado, D.~E. Kaplan, and T.~M. Tait, ``{The Higgs mass bound in
  gauge extensions of the minimal supersymmetric standard model},''
  \href{http://dx.doi.org/10.1088/1126-6708/2004/02/043}{{\em JHEP} {\bfseries
  0402} (2004) 043},
\href{http://arxiv.org/abs/hep-ph/0309149}{{\ttfamily arXiv:hep-ph/0309149
  [hep-ph]}}.

\bibitem{Papucci:2014rja}
M.~Papucci, K.~Sakurai, A.~Weiler, and L.~Zeune, ``{Fastlim: a fast LHC limit
  calculator},''
\href{http://arxiv.org/abs/1402.0492}{{\ttfamily arXiv:1402.0492 [hep-ph]}}.

\bibitem{Kats:2011it}
Y.~Kats and D.~Shih, ``{Light Stop NLSPs at the Tevatron and LHC},''
  \href{http://dx.doi.org/10.1007/JHEP08(2011)049}{{\em JHEP} {\bfseries 1108}
  (2011) 049},
\href{http://arxiv.org/abs/1106.0030}{{\ttfamily arXiv:1106.0030 [hep-ph]}}.

\bibitem{Essig:2011qg}
R.~Essig, E.~Izaguirre, J.~Kaplan, and J.~G. Wacker, ``{Heavy Flavor Simplified
  Models at the LHC},'' \href{http://dx.doi.org/10.1007/JHEP01(2012)074}{{\em
  JHEP} {\bfseries 1201} (2012) 074},
\href{http://arxiv.org/abs/1110.6443}{{\ttfamily arXiv:1110.6443 [hep-ph]}}.

\bibitem{Kats:2011qh}
Y.~Kats, P.~Meade, M.~Reece, and D.~Shih, ``{The Status of GMSB After 1/fb at
  the LHC},'' \href{http://dx.doi.org/10.1007/JHEP02(2012)115}{{\em JHEP}
  {\bfseries 1202} (2012) 115},
\href{http://arxiv.org/abs/1110.6444}{{\ttfamily arXiv:1110.6444 [hep-ph]}}.

\bibitem{Brust:2011tb}
C.~Brust, A.~Katz, S.~Lawrence, and R.~Sundrum, ``{SUSY, the Third Generation
  and the LHC},'' \href{http://dx.doi.org/10.1007/JHEP03(2012)103}{{\em JHEP}
  {\bfseries 1203} (2012) 103},
\href{http://arxiv.org/abs/1110.6670}{{\ttfamily arXiv:1110.6670 [hep-ph]}}.

\bibitem{Papucci:2011wy}
M.~Papucci, J.~T. Ruderman, and A.~Weiler, ``{Natural SUSY Endures},''
  \href{http://dx.doi.org/10.1007/JHEP09(2012)035}{{\em JHEP} {\bfseries 1209}
  (2012) 035},
\href{http://arxiv.org/abs/1110.6926}{{\ttfamily arXiv:1110.6926 [hep-ph]}}.

\bibitem{Evans:2013jna}
J.~A. Evans, Y.~Kats, D.~Shih, and M.~J. Strassler, ``{Toward Full LHC Coverage
  of Natural Supersymmetry},''
  \href{http://dx.doi.org/10.1007/JHEP07(2014)101}{{\em JHEP} {\bfseries 1407}
  (2014) 101},
\href{http://arxiv.org/abs/1310.5758}{{\ttfamily arXiv:1310.5758 [hep-ph]}}.

\bibitem{Kowalska:2013ica}
K.~Kowalska and E.~M. Sessolo, ``{Natural MSSM after the LHC 8 TeV run},''
  \href{http://dx.doi.org/10.1103/PhysRevD.88.075001}{{\em Phys.Rev.}
  {\bfseries D88} (2013) 075001},
\href{http://arxiv.org/abs/1307.5790}{{\ttfamily arXiv:1307.5790 [hep-ph]}}.

\bibitem{Kribs:2013lua}
G.~D. Kribs, A.~Martin, and A.~Menon, ``{Natural Supersymmetry and Implications
  for Higgs physics},''
  \href{http://dx.doi.org/10.1103/PhysRevD.88.035025}{{\em Phys.Rev.}
  {\bfseries D88} (2013) 035025},
\href{http://arxiv.org/abs/1305.1313}{{\ttfamily arXiv:1305.1313 [hep-ph]}}.

\bibitem{Han:2013kga}
C.~Han, K.-i. Hikasa, L.~Wu, J.~M. Yang, and Y.~Zhang, ``{Current experimental
  bounds on stop mass in natural SUSY},''
  \href{http://dx.doi.org/10.1007/JHEP10(2013)216}{{\em JHEP} {\bfseries 1310}
  (2013) 216},
\href{http://arxiv.org/abs/1308.5307}{{\ttfamily arXiv:1308.5307 [hep-ph]}}.

\bibitem{Evans:2012bf}
J.~A. Evans and Y.~Kats, ``{LHC Coverage of RPV MSSM with Light Stops},''
  \href{http://dx.doi.org/10.1007/JHEP04(2013)028}{{\em JHEP} {\bfseries 1304}
  (2013) 028},
\href{http://arxiv.org/abs/1209.0764}{{\ttfamily arXiv:1209.0764 [hep-ph]}}.

\bibitem{Evans:2013uwa}
J.~A. Evans and Y.~Kats, ``{LHC searches examined via the RPV MSSM},''
\href{http://arxiv.org/abs/1311.0890}{{\ttfamily arXiv:1311.0890 [hep-ph]}}.

\bibitem{Pappadopulo:2014qza}
D.~Pappadopulo, A.~Thamm, R.~Torre, and A.~Wulzer, ``{Heavy Vector Triplets:
  Bridging Theory and Data},''
\href{http://arxiv.org/abs/1402.4431}{{\ttfamily arXiv:1402.4431 [hep-ph]}}.

\bibitem{Berger:2012ec}
J.~Berger, J.~Hubisz, and M.~Perelstein, ``{A Fermionic Top Partner:
  Naturalness and the LHC},''
  \href{http://dx.doi.org/10.1007/JHEP07(2012)016}{{\em JHEP} {\bfseries 1207}
  (2012) 016},
\href{http://arxiv.org/abs/1205.0013}{{\ttfamily arXiv:1205.0013 [hep-ph]}}.

\bibitem{Cacciapaglia:2012dd}
G.~Cacciapaglia, A.~Deandrea, L.~Panizzi, S.~Perries, and V.~Sordini, ``{Heavy
  Vector-like quark with charge 5/3 at the LHC},''
  \href{http://dx.doi.org/10.1007/JHEP03(2013)004}{{\em JHEP} {\bfseries 1303}
  (2013) 004},
\href{http://arxiv.org/abs/1211.4034}{{\ttfamily arXiv:1211.4034 [hep-ph]}}.

\bibitem{Azatov:2013hya}
A.~Azatov, M.~Salvarezza, M.~Son, and M.~Spannowsky, ``{Boosting Top Partner
  Searches in Composite Higgs Models},''
  \href{http://dx.doi.org/10.1103/PhysRevD.89.075001}{{\em Phys.Rev.}
  {\bfseries D89} (2014) 075001},
\href{http://arxiv.org/abs/1308.6601}{{\ttfamily arXiv:1308.6601 [hep-ph]}}.

\bibitem{Reuter:2013iya}
J.~Reuter, M.~Tonini, and M.~de~Vries, ``{Littlest Higgs with T-parity: Status
  and Prospects},'' \href{http://dx.doi.org/10.1007/JHEP02(2014)053}{{\em JHEP}
  {\bfseries 1402} (2014) 053},
\href{http://arxiv.org/abs/1310.2918}{{\ttfamily arXiv:1310.2918 [hep-ph]}}.

\bibitem{Gillioz:2013pba}
M.~Gillioz, R.~Gršber, A.~Kapuvari, and M.~MŸhlleitner, ``{Vector-like Bottom
  Quarks in Composite Higgs Models},''
  \href{http://dx.doi.org/10.1007/JHEP03(2014)037}{{\em JHEP} {\bfseries 1403}
  (2014) 037},
\href{http://arxiv.org/abs/1311.4453}{{\ttfamily arXiv:1311.4453 [hep-ph]}}.

\bibitem{Matsedonskyi:2014lla}
O.~Matsedonskyi, F.~Riva, and T.~Vantalon, ``{Composite Charge 8/3 Resonances
  at the LHC},'' \href{http://dx.doi.org/10.1007/JHEP04(2014)059}{{\em JHEP}
  {\bfseries 1404} (2014) 059},
\href{http://arxiv.org/abs/1401.3740}{{\ttfamily arXiv:1401.3740 [hep-ph]}}.

\bibitem{Fan:2011yu}
J.~Fan, M.~Reece, and J.~T. Ruderman, ``{Stealth Supersymmetry},''
  \href{http://dx.doi.org/10.1007/JHEP11(2011)012}{{\em JHEP} {\bfseries 1111}
  (2011) 012},
\href{http://arxiv.org/abs/1105.5135}{{\ttfamily arXiv:1105.5135 [hep-ph]}}.

\bibitem{Murayama:2012jh}
H.~Murayama, Y.~Nomura, S.~Shirai, and K.~Tobioka, ``{Compact Supersymmetry},''
  \href{http://dx.doi.org/10.1103/PhysRevD.86.115014}{{\em Phys.Rev.}
  {\bfseries D86} (2012) 115014},
\href{http://arxiv.org/abs/1206.4993}{{\ttfamily arXiv:1206.4993 [hep-ph]}}.

\bibitem{Bharucha:2013epa}
A.~Bharucha, S.~Heinemeyer, and F.~Pahlen, ``{Direct Chargino-Neutralino
  Production at the LHC: Interpreting the Exclusion Limits in the Complex
  MSSM},'' \href{http://dx.doi.org/10.1140/epjc/s10052-013-2629-x}{{\em
  Eur.Phys.J.} {\bfseries C73} (2013) 2629},
\href{http://arxiv.org/abs/1307.4237}{{\ttfamily arXiv:1307.4237}}.

\bibitem{Schwaller:2013baa}
P.~Schwaller and J.~Zurita, ``{Compressed electroweakino spectra at the LHC},''
  \href{http://dx.doi.org/10.1007/JHEP03(2014)060}{{\em JHEP} {\bfseries 1403}
  (2014) 060},
\href{http://arxiv.org/abs/1312.7350}{{\ttfamily arXiv:1312.7350 [hep-ph]}}.

\bibitem{Baer:2014cua}
H.~Baer, A.~Mustafayev, and X.~Tata, ``{Monojets and mono-photons from light
  higgsino pair production at LHC14},''
  \href{http://dx.doi.org/10.1103/PhysRevD.89.055007}{{\em Phys.Rev.}
  {\bfseries D89} (2014) 055007},
\href{http://arxiv.org/abs/1401.1162}{{\ttfamily arXiv:1401.1162 [hep-ph]}}.

\bibitem{Han:2014kaa}
Z.~Han, G.~D. Kribs, A.~Martin, and A.~Menon, ``{Hunting quasidegenerate
  Higgsinos},'' \href{http://dx.doi.org/10.1103/PhysRevD.89.075007}{{\em
  Phys.Rev.} {\bfseries D89} (2014) 075007},
\href{http://arxiv.org/abs/1401.1235}{{\ttfamily arXiv:1401.1235 [hep-ph]}}.

\bibitem{Buchmueller:2013exa}
O.~Buchmueller and J.~Marrouche, ``{Universal mass limits on gluino and
  third-generation squarks in the context of Natural-like SUSY spectra},''
  \href{http://dx.doi.org/10.1142/S0217751X14500328}{{\em Int.J.Mod.Phys.}
  {\bfseries A29} (2014) 1450032},
\href{http://arxiv.org/abs/1304.2185}{{\ttfamily arXiv:1304.2185 [hep-ph]}}.

\bibitem{Barnard:2014joa}
J.~Barnard and B.~Farmer, ``{A simple technique for combining simplified models
  and its application to direct stop production},''
  \href{http://dx.doi.org/10.1007/JHEP06(2014)132}{{\em JHEP} {\bfseries 1406}
  (2014) 132},
\href{http://arxiv.org/abs/1402.3298}{{\ttfamily arXiv:1402.3298 [hep-ph]}}.

\bibitem{Howe:2012xe}
K.~Howe and P.~Saraswat, ``{Excess Higgs Production in Neutralino Decays},''
  \href{http://dx.doi.org/10.1007/JHEP10(2012)065}{{\em JHEP} {\bfseries 1210}
  (2012) 065},
\href{http://arxiv.org/abs/1208.1542}{{\ttfamily arXiv:1208.1542 [hep-ph]}}.

\bibitem{Arbey:2012fa}
A.~Arbey, M.~Battaglia, and F.~Mahmoudi, ``{Higgs Production in Neutralino
  Decays in the MSSM - The LHC and a Future e+e- Collider},''
\href{http://arxiv.org/abs/1212.6865}{{\ttfamily arXiv:1212.6865 [hep-ph]}}.

\bibitem{Blanke:2013uia}
M.~Blanke, G.~F. Giudice, P.~Paradisi, G.~Perez, and J.~Zupan, ``{Flavoured
  Naturalness},'' \href{http://dx.doi.org/10.1007/JHEP06(2013)022}{{\em JHEP}
  {\bfseries 1306} (2013) 022},
\href{http://arxiv.org/abs/1302.7232}{{\ttfamily arXiv:1302.7232 [hep-ph]}}.

\bibitem{Mahbubani:2012qq}
R.~Mahbubani, M.~Papucci, G.~Perez, J.~T. Ruderman, and A.~Weiler, ``{Light
  Nondegenerate Squarks at the LHC},''
  \href{http://dx.doi.org/10.1103/PhysRevLett.110.151804}{{\em Phys.Rev.Lett.}
  {\bfseries 110} no.~15, (2013) 151804},
\href{http://arxiv.org/abs/1212.3328}{{\ttfamily arXiv:1212.3328 [hep-ph]}}.

\bibitem{Heikinheimo:2011fk}
M.~Heikinheimo, M.~Kellerstein, and V.~Sanz, ``{How Many Supersymmetries?},''
  \href{http://dx.doi.org/10.1007/JHEP04(2012)043}{{\em JHEP} {\bfseries 1204}
  (2012) 043},
\href{http://arxiv.org/abs/1111.4322}{{\ttfamily arXiv:1111.4322 [hep-ph]}}.

\bibitem{Kribs:2012gx}
G.~D. Kribs and A.~Martin, ``{Supersoft Supersymmetry is Super-Safe},''
  \href{http://dx.doi.org/10.1103/PhysRevD.85.115014}{{\em Phys.Rev.}
  {\bfseries D85} (2012) 115014},
\href{http://arxiv.org/abs/1203.4821}{{\ttfamily arXiv:1203.4821 [hep-ph]}}.

\bibitem{Burdman:2006tz}
G.~Burdman, Z.~Chacko, H.-S. Goh, and R.~Harnik, ``{Folded supersymmetry and
  the LEP paradox},''
  \href{http://dx.doi.org/10.1088/1126-6708/2007/02/009}{{\em JHEP} {\bfseries
  0702} (2007) 009},
\href{http://arxiv.org/abs/hep-ph/0609152}{{\ttfamily arXiv:hep-ph/0609152
  [hep-ph]}}.

\bibitem{Falkowski:2006qq}
A.~Falkowski, S.~Pokorski, and M.~Schmaltz, ``{Twin SUSY},''
  \href{http://dx.doi.org/10.1103/PhysRevD.74.035003}{{\em Phys.Rev.}
  {\bfseries D74} (2006) 035003},
\href{http://arxiv.org/abs/hep-ph/0604066}{{\ttfamily arXiv:hep-ph/0604066
  [hep-ph]}}.

\bibitem{Chang:2006ra}
S.~Chang, L.~J. Hall, and N.~Weiner, ``{A Supersymmetric twin Higgs},''
  \href{http://dx.doi.org/10.1103/PhysRevD.75.035009}{{\em Phys.Rev.}
  {\bfseries D75} (2007) 035009},
\href{http://arxiv.org/abs/hep-ph/0604076}{{\ttfamily arXiv:hep-ph/0604076
  [hep-ph]}}.

\bibitem{CahillRowley:2012kx}
M.~W. Cahill-Rowley, J.~L. Hewett, A.~Ismail, and T.~G. Rizzo, ``{More Energy,
  More Searches, but the pMSSM Lives On},''
  \href{http://dx.doi.org/10.1103/PhysRevD.88.035002}{{\em Phys.Rev.}
  {\bfseries D88} (2013) 035002},
\href{http://arxiv.org/abs/1211.1981}{{\ttfamily arXiv:1211.1981 [hep-ph]}}.

\bibitem{CahillRowley:2012rv}
M.~W. Cahill-Rowley, J.~L. Hewett, A.~Ismail, and T.~G. Rizzo, ``{The Higgs
  Sector and Fine-Tuning in the pMSSM},''
  \href{http://dx.doi.org/10.1103/PhysRevD.86.075015}{{\em Phys.Rev.}
  {\bfseries D86} (2012) 075015},
\href{http://arxiv.org/abs/1206.5800}{{\ttfamily arXiv:1206.5800 [hep-ph]}}.

\bibitem{CahillRowley:2012cb}
M.~W. Cahill-Rowley, J.~L. Hewett, S.~Hoeche, A.~Ismail, and T.~G. Rizzo,
  ``{The New Look pMSSM with Neutralino and Gravitino LSPs},''
  \href{http://dx.doi.org/10.1140/epjc/s10052-012-2156-1}{{\em Eur.Phys.J.}
  {\bfseries C72} (2012) 2156},
\href{http://arxiv.org/abs/1206.4321}{{\ttfamily arXiv:1206.4321 [hep-ph]}}.

\bibitem{Conley:2011nn}
J.~A. Conley, J.~S. Gainer, J.~L. Hewett, M.~P. Le, and T.~G. Rizzo,
  ``{Supersymmetry Without Prejudice at the 7 TeV LHC},'' {\em Physical Review
  D} (2011) ,
\href{http://arxiv.org/abs/1103.1697}{{\ttfamily arXiv:1103.1697 [hep-ph]}}.

\bibitem{Sekmen:2011cz}
S.~Sekmen, S.~Kraml, J.~Lykken, F.~Moortgat, S.~Padhi, {\em et~al.},
  ``{Interpreting LHC SUSY searches in the phenomenological MSSM},''
  \href{http://dx.doi.org/10.1007/JHEP02(2012)075}{{\em JHEP} {\bfseries 1202}
  (2012) 075},
\href{http://arxiv.org/abs/1109.5119}{{\ttfamily arXiv:1109.5119 [hep-ph]}}.

\bibitem{Carena:2012he}
M.~Carena, J.~Lykken, S.~Sekmen, N.~R. Shah, and C.~E. Wagner, ``{The pMSSM
  Interpretation of LHC Results Using Rernormalization Group Invariants},''
  \href{http://dx.doi.org/10.1103/PhysRevD.86.075025}{{\em Phys.Rev.}
  {\bfseries D86} (2012) 075025},
\href{http://arxiv.org/abs/1205.5903}{{\ttfamily arXiv:1205.5903 [hep-ph]}}.

\bibitem{Meade:2010ji}
P.~Meade, M.~Reece, and D.~Shih, ``{Long-Lived Neutralino NLSPs},''
  \href{http://dx.doi.org/10.1007/JHEP10(2010)067}{{\em JHEP} {\bfseries 1010}
  (2010) 067},
\href{http://arxiv.org/abs/1006.4575}{{\ttfamily arXiv:1006.4575 [hep-ph]}}.

\bibitem{Graham:2012th}
P.~W. Graham, D.~E. Kaplan, S.~Rajendran, and P.~Saraswat, ``{Displaced
  Supersymmetry},'' \href{http://dx.doi.org/10.1007/JHEP07(2012)149}{{\em JHEP}
  {\bfseries 1207} (2012) 149},
\href{http://arxiv.org/abs/1204.6038}{{\ttfamily arXiv:1204.6038 [hep-ph]}}.

\bibitem{Zurek:2010xf}
K.~M. Zurek, ``{TASI 2009 Lectures: Searching for Unexpected Physics at the
  LHC},''
\href{http://arxiv.org/abs/1001.2563}{{\ttfamily arXiv:1001.2563 [hep-ph]}}.

\bibitem{Strassler:2006qa}
M.~J. Strassler, ``{Possible effects of a hidden valley on supersymmetric
  phenomenology},''
\href{http://arxiv.org/abs/hep-ph/0607160}{{\ttfamily arXiv:hep-ph/0607160
  [hep-ph]}}.

\bibitem{ArkaniHamed:2008qn}
N.~Arkani-Hamed, D.~P. Finkbeiner, T.~R. Slatyer, and N.~Weiner, ``{A Theory of
  Dark Matter},'' \href{http://dx.doi.org/10.1103/PhysRevD.79.015014}{{\em
  Phys.Rev.} {\bfseries D79} (2009) 015014},
\href{http://arxiv.org/abs/0810.0713}{{\ttfamily arXiv:0810.0713 [hep-ph]}}.

\bibitem{ArkaniHamed:2008qp}
N.~Arkani-Hamed and N.~Weiner, ``{LHC Signals for a SuperUnified Theory of Dark
  Matter},'' \href{http://dx.doi.org/10.1088/1126-6708/2008/12/104}{{\em JHEP}
  {\bfseries 0812} (2008) 104},
\href{http://arxiv.org/abs/0810.0714}{{\ttfamily arXiv:0810.0714 [hep-ph]}}.

\bibitem{Aad:2012qua}
{\bfseries ATLAS} Collaboration, G.~Aad {\em et~al.}, ``{A search for prompt
  lepton-jets in $pp$ collisions at $\sqrt{s}=7$ TeV with the ATLAS
  detector},'' \href{http://dx.doi.org/10.1016/j.physletb.2013.01.034}{{\em
  Phys.Lett.} {\bfseries B719} (2013) 299--317},
\href{http://arxiv.org/abs/1212.5409}{{\ttfamily arXiv:1212.5409}}.

\bibitem{Chatrchyan:2012cg}
{\bfseries CMS} Collaboration, S.~Chatrchyan {\em et~al.}, ``{Search for a
  non-standard-model Higgs boson decaying to a pair of new light bosons in
  four-muon final states},''
  \href{http://dx.doi.org/10.1016/j.physletb.2013.09.009}{{\em Phys.Lett.}
  {\bfseries B726} (2013) 564--586},
\href{http://arxiv.org/abs/1210.7619}{{\ttfamily arXiv:1210.7619 [hep-ex]}}.

\bibitem{Kang:2008ea}
J.~Kang and M.~A. Luty, ``{Macroscopic Strings and 'Quirks' at Colliders},''
  \href{http://dx.doi.org/10.1088/1126-6708/2009/11/065}{{\em JHEP} {\bfseries
  0911} (2009) 065},
\href{http://arxiv.org/abs/0805.4642}{{\ttfamily arXiv:0805.4642 [hep-ph]}}.

\bibitem{CMS:2012xi}
{\bfseries CMS} Collaboration, S.~Chatrchyan {\em et~al.}, ``{Search for
  fractionally charged particles in $pp$ collisions at $\sqrt{s}=7$ TeV},''
  \href{http://dx.doi.org/10.1103/PhysRevD.87.092008}{{\em Phys.Rev.}
  {\bfseries D87} no.~9, (2013) 092008},
\href{http://arxiv.org/abs/1210.2311}{{\ttfamily arXiv:1210.2311 [hep-ex]}}.

\bibitem{EuropeanStrategyforParticlePhysicsPreparatoryGroup:2013fia}
{\bfseries European Strategy for Particle Physics Preparatory Group}
  Collaboration, R.~Aleksan {\em et~al.}, ``{Physics Briefing Book: Input for
  the Strategy Group to draft the update of the European Strategy for Particle
  Physics},''.
CERN-ESG-005.

\bibitem{Rosner:2014pja}
J.~Rosner, M.~Bardeen, W.~Barletta, L.~Bauerdick, R.~Bernstein, {\em et~al.},
  ``{Planning the Future of U.S. Particle Physics (Snowmass 2013): Chapter 1:
  Summary},''
\href{http://arxiv.org/abs/1401.6075}{{\ttfamily arXiv:1401.6075 [hep-ex]}}.

\bibitem{ECFA-13-284}
``{ECFA High Luminosity LHC Experiments Workshop: Physics and Technology
  Challenges. 94th Plenary ECFA meeting},''. ECFA-13-284.

\bibitem{ATLAS:2013hta}
{\bfseries ATLAS} Collaboration, ``{Physics at a High-Luminosity LHC with
  ATLAS},''
\href{http://arxiv.org/abs/1307.7292}{{\ttfamily arXiv:1307.7292 [hep-ex]}}.

\bibitem{CMS:2013xfa}
{\bfseries CMS} Collaboration, ``{Projected Performance of an Upgraded CMS
  Detector at the LHC and HL-LHC: Contribution to the Snowmass Process},''
\href{http://arxiv.org/abs/1307.7135}{{\ttfamily arXiv:1307.7135}}.

\bibitem{ATLAS-PHYS-PUB-2013-011}
``{Prospects for benchmark Supersymmetry searches at the high luminosity LHC
  with the ATLAS Detector},''. ATL-PHYS-PUB-2013-011.

\bibitem{Han:2013usa}
C.~Han, A.~Kobakhidze, N.~Liu, A.~Saavedra, L.~Wu, {\em et~al.}, ``{Probing
  Light Higgsinos in Natural SUSY from Monojet Signals at the LHC},''
  \href{http://dx.doi.org/10.1007/JHEP02(2014)049}{{\em JHEP} {\bfseries 1402}
  (2014) 049},
\href{http://arxiv.org/abs/1310.4274}{{\ttfamily arXiv:1310.4274 [hep-ph]}}.

\bibitem{Bhattacherjee:2013wna}
B.~Bhattacherjee, A.~Choudhury, K.~Ghosh, and S.~Poddar, ``{Compressed
  supersymmetry at 14 TeV LHC},''
  \href{http://dx.doi.org/10.1103/PhysRevD.89.037702}{{\em Phys.Rev.}
  {\bfseries D89} no.~3, (2014) 037702},
\href{http://arxiv.org/abs/1308.1526}{{\ttfamily arXiv:1308.1526 [hep-ph]}}.

\bibitem{Delannoy:2013ata}
A.~G. Delannoy, B.~Dutta, A.~Gurrola, W.~Johns, T.~Kamon, {\em et~al.},
  ``{Probing Dark Matter at the LHC using Vector Boson Fusion Processes},''
  \href{http://dx.doi.org/10.1103/PhysRevLett.111.061801}{{\em Phys.Rev.Lett.}
  {\bfseries 111} (2013) 061801},
\href{http://arxiv.org/abs/1304.7779}{{\ttfamily arXiv:1304.7779 [hep-ph]}}.

\bibitem{Weinberg:2012es}
D.~H. Weinberg, M.~J. Mortonson, D.~J. Eisenstein, C.~Hirata, A.~G. Riess, {\em
  et~al.}, ``{Observational Probes of Cosmic Acceleration},''
  \href{http://dx.doi.org/10.1016/j.physrep.2013.05.001}{{\em Phys.Rept.}
  {\bfseries 530} (2013) 87--255},
\href{http://arxiv.org/abs/1201.2434}{{\ttfamily arXiv:1201.2434
  [astro-ph.CO]}}.

\bibitem{Weinberg:1988cp}
S.~Weinberg, ``{The Cosmological Constant Problem},''
\href{http://dx.doi.org/10.1103/RevModPhys.61.1}{{\em Rev.Mod.Phys.} {\bfseries
  61} (1989) 1--23}.

\end{thebibliography}\endgroup

\end{document}